\documentclass[11pt,a4paper]{article}    
\usepackage{jheppub}
\makeatletter

\usepackage{fontenc}
\usepackage[utf8]{inputenc}
\usepackage{float}
\usepackage{units}
\usepackage{amsmath}
\usepackage{amssymb}
\usepackage{graphicx}
\usepackage{color}
\usepackage{esint}
\usepackage{youngtab}

\usepackage{nicematrix}
\usepackage{bbold}

\hypersetup{%
	pdftitle   = {Chiral Carroll string near Schwarzschild black hole},
	pdfkeywords = {Carrollian strings, black holes, non-Lorentzian geometry},
	pdfauthor  = {Kedar S.~Kolekar},
}

\usepackage{wrapfig}
\usepackage{slashed}
\usepackage{epsfig}
\usepackage{tensor}
\def\be{\begin{eqnarray}}
\def\ee{\end{eqnarray}}

\def\Dslash{\,\,{\raise.15ex\hbox{/}\mkern-12mu D}}
\def\Dbarslash{\,\,{\raise.15ex\hbox{/}\mkern-12mu {\bar D}}}
\def\delslash{\,\,{\raise.15ex\hbox{/}\mkern-9mu \partial}}
\def\delbarslash{\,\,{\raise.15ex\hbox{/}\mkern-9mu {\bar\partial}}}
\def\pslash{\,\,{\raise.15ex\hbox{/}\mkern-9mu p}}
\def\calDslash{\,\,{\raise.15ex\hbox{/}\mkern-12mu {\cal D}}}

\newcommand{\calO}{{\cal O}}

\newcommand{\zero}{\mbox{\tiny $0$}}

\newcommand{\half}{\frac{1}{2}}

\newcommand{\eff}{\text{eff}}

\def\ie{\emph{i.e.}}

\usepackage{scalerel}
\newcommand{\ost}[2]{\overset{\makebox[0pt]{$\scriptscriptstyle{\scaleto{(\hspace{-0.06em}#2\hspace{-0.06em})}{5pt}}$}}{#1}{}}

\def\implies{\Rightarrow}

\def\lae{\mathrel{\mathop{\smash{\lower .5 ex \hbox{$\stackrel<\sim$}}}}}
\def\lae{\mathrel{\mathop{\smash{\lower .5 ex \hbox{$\stackrel>\sim$}}}}}

\usepackage[useregional]{datetime2}

\usepackage{float}
\usepackage{tikz}
\usepackage{pgfplots}
\pgfplotsset{compat=1.16}
\usetikzlibrary{calc,arrows,cd}
\usetikzlibrary{shapes,decorations}
\usepackage{makecell}
\pgfdeclarelayer{nodelayer}
\pgfdeclarelayer{edgelayer}
\pgfsetlayers{edgelayer,nodelayer,main}
\tikzstyle{ghost}=[fill={rgb,255: red,140; green,76; blue,150}, draw=black, shape=circle,scale=0.6]
\tikzstyle{real_ghost}=[fill=none, draw=none, shape=circle]
\tikzstyle{none}=[fill=none, draw=none, shape=circle]
\tikzstyle{Red_Circle}=[fill=none, draw=red, shape=circle]

\tikzstyle{BlackLine}=[-, draw=black, fill=white]
\tikzstyle{Arrow}=[<-, thick]
\tikzstyle{thin_black_line}=[-, fill=blue]
\tikzstyle{thin_black_line_red}=[-, fill=red]
\tikzstyle{thin_black_line_semi_purple}=[-, fill={rgb,255: red,1; green,1; blue,255}]
\tikzstyle{thin_black_line_purple}=[-, fill={rgb,255: red,200; green,1; blue,255}]
\tikzstyle{thin_black_line_turquoise}=[-, fill={rgb,255: red,1; green,200; blue,200}]
\tikzstyle{thin_black_line_gray}=[-, fill=gray]
\tikzstyle{thin_black_line_null}=[-, fill=orange]
\tikzstyle{Double_Arrow}=[<->]
\tikzstyle{Dashed_arrow}=[->, dashed, draw=red]
\tikzstyle{Dashed_arrow_gray}=[->, dashed, draw=gray]
\tikzstyle{BlackLine_dash}=[-, draw=red, fill=white, dashed]
\tikzstyle{Arrow_order}=[<-, draw=red]

\DeclareMathAlphabet{\mathdutchcal}{U}{dutchcal}{m}{n}
\SetMathAlphabet{\mathdutchcal}{bold}{U}{dutchcal}{b}{n}
\DeclareMathAlphabet{\mathdutchbcal}{U}{dutchcal}{b}{n}
\title{Chiral Carroll string near Schwarzschild black hole}

\author[a]{Kedar S.~Kolekar}
\affiliation[a]{Beijing Institute of Mathematical Sciences and Applications, Beijing 101408, China.}

\emailAdd{kedar@bimsa.cn}

\abstract{We obtain the action for the chiral Carroll string moving in a curved string-Carroll target spacetime from a Carroll expansion, in powers of the speed of light $c$, of the action for the relativistic closed bosonic string.
Applying this setup to a classical closed bosonic string approaching the $4$-dimensional Schwarzschild black hole, we study the dynamics of the string when it behaves as a chiral Carroll string with the near-horizon string-Carroll expanded geometry as the target spacetime. We find that the chiral Carroll string can have either left-moving or right-moving fluctuations on the horizon, while exhibiting a non-trivial motion in the Rindler spacetime which depends on these horizon fluctuations.}

\usepackage{todonotes}
\usepackage{mathrsfs}
\usepackage{amscd}

\usepackage{amsthm}

\theoremstyle{remark}

\usepackage{tcolorbox}
\usepackage{cancel}
\usepackage{xcolor}

\usepackage{tikz}

\DeclareFontFamily{U}{skulls}{}
\DeclareFontShape{U}{skulls}{m}{n}{ <-> skull }{}

\begin{document}
\pagestyle{plain} \setcounter{page}{1}
\newcounter{bean}
\baselineskip16pt \setcounter{section}{0}
\maketitle
\flushbottom

%%%%%%%%%%%%%%%%%%%%%%%%%%%%%%
%%%%% Sec-1: Introduction %%%%%%%%%%%%%%
\newpage

\section{Introduction}
\label{sec:intro}

One of the useful approaches to understand the properties of black holes is to send probes into black holes. In the framework of string theory, one sends a string towards a black hole and studies how the black hole influences the dynamics of the infalling sting.
%Understanding black holes is an important problem in theoretical physics. It can be done through many approaches; one of them being sending probes to the black holes. In the framework of string theory, we can send strings as probes and study how the black hole influences the dynamics of an infalling string.
In \cite{Bagchi:2023cfp, Bagchi:2024rje} (see also \cite{Banerjee:2025bkg}), it was found that an infalling closed bosonic string approaching the horizon of a non-extremal black hole perceives the near-horizon region as a string-Carroll expanded geometry, thus behaving as a Carroll string. Carroll strings refer to strings moving in target spacetimes having Carroll structures while the worldsheet can be Lorentzian or Carrollian. Carroll strings were obtained from different ways of taking the Carroll limit in \cite{Cardona:2016ytk,Blair:2023noj,Gomis:2023eav,Bagchi:2023cfp,Harksen:2024bnh,Casalbuoni:2024jmj,Bagchi:2024rje}, and their quantisation was discussed in \cite{Figueroa-OFarrill:2025njv}. On the other hand, the string having a Carrollian worldsheet embedded in a Lorentzian target spacetime is referred to as the tensionless null string \cite{Isberg:1993av,Bagchi:2013bga,Bagchi:2015nca} (see \cite{Bagchi:2026wcu} for a recent review, and references therein). Carroll theories form a class of non-Lorentzian theories that arise in the Carroll limit $c\rightarrow 0$, where $c$ is the speed of light, of relativistic theories \cite{LBLL,NDS,Henneaux:1979vn,Duval:2014uoa}. Some of the recent reviews \cite{Bergshoeff:2022eog,Bagchi:2025vri,Ciambelli:2025unn,Ruzziconi:2026bix} discuss multiple aspects of Carroll physics, and contain an extensive list of references.

%On the other hand, the opposite limit $c\rightarrow\infty$ is the well known Galilean limit which gives rise to non-relativistic regime of relativistic theories \cite{}. Non-relativistic string theories for strings embedded in Newton-Cartan geometries as target spacetimes are discussed in \cite{} (see also the recent review \cite{}).

Two kinds of Carroll strings are discussed in \cite{Bagchi:2024rje}:~an \emph{electric} Carroll string having a Carrollian worldsheet and a \emph{magnetic} Carroll string having a Lorentzian worldsheet, both embedded in a string-Carroll target spacetime. String-Carroll spacetime is a generalisation of the Carroll spacetime that now has a two-dimensional null subspace referred to as the longitudinal space, fibred over a Riemannian base space referred to as the transverse space.
%whereas Carroll geometry has a single null direction.
The electric and magnetic Carroll strings are obtained by doing two inequivalent scalings of the string tension and the Lagrange multipliers that impose the Hamiltonian constraints for the string when taking the Carroll limit $c\rightarrow 0$ (or doing a small-$c$ expansion) of the relativistic string. Apart from the worldsheet geometry, another distinguishing feature between the two strings lies in the dynamics of the transverse embedding fields:~they are constant for the magnetic Carroll string while exhibiting non-trivial dynamics for the electric Carroll string. In particular, for a string approaching the horizon of a $4$-dimensional Schwarzschild black hole, the magnetic Carroll string freezes on the transverse sphere, \ie~the horizon, whereas the electric string extends along the horizon and can wrap around it.

A third kind of Carroll string named the \emph{chiral Carroll string} was introduced in \cite{Harksen:2024bnh}. It was obtained from an appropriately defined Carroll limit, as $\alpha'\rightarrow 0$, of the Polyakov action for a relativistic closed bosonic string embedded in a Minkowski target spacetime.
%and asymmetrically rescaling the longitudinal and transverse embedding fields with $\alpha'$.
The resulting chiral Carroll string has a Lorentzian worldsheet, like the magnetic Carroll string, and is embedded in a flat string-Carroll target spacetime. However, unlike the magnetic Carroll string, it exhibits non-trivial dynamics for the transverse embedding fields by allowing them to be functions of either the left-moving or the right-moving coordinate on the worldsheet.

In this paper, we generalise the definition of the chiral Carroll string to a curved string-Carroll target spacetime. In particular, we employ the framework of \cite{Bagchi:2024rje} and obtain the action for the chiral Carroll string at leading order in a Carroll expansion, in powers of $c^2$, of the relativistic string action. We also write down the next-to-leading order action in this expansion that governs the dynamics of the sub-leading embedding fields in the chiral Carroll string sector. Then, using this setup, we analyse the motion of an infalling classical closed bosonic string in the near-horizon region of the Schwarzschild black hole, when it behaves as the chiral Carroll string embedded in the Rindler $\times$ $S^2$ string-Carroll target spacetime.
%the leading term in the near-horizon expansion of the Schwarzschild geometry.
We find that the chiral Carroll string has either left-moving or right-moving fluctuations on the horizon ($S^2$), while having non-trivial motion in the $2$-dimensional Rindler spacetime.

This paper is organised as follows. In section~\ref{sec:string-Carroll expansion}, we review the string-Carroll expansion of Lorentzian spacetimes in powers of the speed of light, $c$. In section~\ref{sec:magnetic string}, we provide an alternate description of the magnetic Carroll string, which also provides a set up to formulate the chiral Carroll string. In section~\ref{sec:chiral string from expansion}, we obtain the Lagrangians for the chiral Carroll string at the leading order (LO) and the next-to-leading order (NLO) from the string-Carroll expansion of the Lagrangian for the relativistic closed bosonic string. In section~\ref{sec:chiral string near black hole}, we study the dynamics of a closed bosonic string approaching the Schwarzschild black hole, when it behaves as the chiral Carroll string. In section~\ref{sec:discussion}, we summarise the paper and comment on future directions. In appendix~\ref{app:magnetic string near black hole}, we verify the equivalence of the two descriptions of the magnetic Carroll string moving in the near-horizon region of the Schwarzschild black hole.

%%%%%%%%%%%%%%%%%%%%%%%%%%%%%
%%% Sec-2: String-Carroll expansion %%%%%%%%%

\section{Review:~String-Carroll expansion}\label{sec:string-Carroll expansion}

In this section, we briefly review the string-Carroll geometry and the string-Carroll expansion of Lorentzian spacetimes, introduced in \cite{Bagchi:2023cfp,Bagchi:2024rje}. A detailed discussion including the causal structure of string-Carroll spacetimes, string-Carroll symmetries and the string-Carroll expansion of Lorentzian symmetries can be found in \cite{Bagchi:2024rje}.
%\textcolor{red}{For a general Carroll expansion of Einstein's gravity, see \cite{Hansen:2021fxi} (see also \cite{Bhattacharya:2026sla}).}

A curved string-Carroll spacetime in $D$ dimensions is described by degenerate metrics $h_{\mu\nu}$ and $\tau^{\mu\nu}$ of signature $(0,0,\mathbf{1}_{D-2})$ and $(-1,1,\mathbf{0}_{D-2})$, respectively, satisfying the degeneracy relation $h_{\mu\lambda}\tau^{\lambda\nu} = 0$.
%\textcolor{red}{It is a generalisation of the Carroll spacetime, which now has a $2$ dimensional null subspace described by $\tau^{\mu\nu}$ generalising the $1$ dimensional null direction described by the null vector $\tau^{\mu}$.}
The components $\tau^{ab}$ for $a,b=0,1$ give a non-degenerate metric on the $2$-dimensional Minkowski subspace referred to as the longitudinal space, and the components $h_{ij}$ for $i,j = 2,\dots,D-1$ give a non-degenerate metric on the $D-2$ dimensional spatial subspace referred to as the transverse space. A simple example is the flat string-Carroll spacetime described by $h_{\mu\nu} = diag(0,0,\mathbf{1}_{D-2})$ and $\tau^{\mu\nu} = daig(-1,1,\mathbf{0}_{D-2})$.

Carroll geometries are naturally defined in Cartan formalism in terms of the vielbeine in the tangent space of the manifold \cite{Hartong:2015xda,Bergshoeff:2017btm}. For the string-Carroll geometry, the metrics $\tau^{\mu\nu}$ and $h_{\mu\nu}$ are written in terms of the longitudinal vielbeine $\tau^{\mu}_A$ and the spatial vielbeine $e^{A'}_{\mu}$ as
\begin{equation}
	\tau^{\mu\nu} = \tau^{\mu}_A\tau^{\nu}_B\eta^{AB}, \quad h_{\mu\nu} = e^{A'}_{\mu}e^{B'}_{\nu}\delta_{A'B'},
\end{equation}
where $\eta^{AB} = diag(-1,1)$ for $A,B=0,1$ is the Minkowski metric in the longitudinal tangent space and $\delta_{A'B'}$ for $A',B'=2,\dots,D-1$ is the Euclidean metric in the transverse tangent space.

We obtain the string-Carroll metrics from a Carroll expansion in two steps. We first write a pseudo-Riemannian metric $g_{\mu\nu}$ and its inverse $g^{\mu\nu}$ in terms of the Lorentzian vielbeine $E^{\hat{A}}_{\mu}$ and $E_{\hat{A}}^{\mu}$,
\begin{equation}
	g_{\mu\nu} = E_{\mu}^{\hat{A}}E_{\nu}^{\hat{B}}\eta_{\hat{A}\hat{B}}, \quad g^{\mu\nu} = E^{\mu}_{\hat{A}}E^{\nu}_{\hat{B}}\eta^{\hat{A}\hat{B}},
\end{equation}
where $\hat{A} = 0,1,\dots,D-1$ and $\eta_{\hat{A}\hat{B}}$ is the Minkowski metric in the Lorentzian tangent space. The metric $g_{\mu\nu}$ here is a generalisation of the usual pseudo-Riemannian metric, which has two directions scaled by the $c^2$ factor, \ie~the time direction and one of the spatial directions, say $x^1$. Decomposing the Lorentzian vielbeine into longitudinal ($\hat{A} = A$) and transverse ($\hat{A}=A'$) sectors
\begin{equation}
	E^{\mu}_A = \frac{T^{\mu}_A}{c^2}, \quad E_{\mu}^A = c^2 T_{\mu}^A, \quad E^{\mu}_{A'} = E^{\mu}_{A'}, \quad E_{\mu}^{A'} = E_{\mu}^{A'},
\end{equation}
we can write the metric as
\begin{subequations}\label{eq:pul csq metric}
\begin{eqnarray}
	g_{\mu\nu} &=& c^2 T_{\mu}^A T_{\nu}^B\eta_{AB} + E_{\mu}^{A'} E_{\nu}^{B'} \delta_{A'B'} \equiv c^2 T_{\mu\nu} + \Pi_{\mu\nu}, \\
	g^{\mu\nu} &=& \frac{1}{c^2}T^{\mu}_AT^{\nu}_B\eta^{AB} + E^{\mu}_{A'}E^{\nu}_{B'}\delta^{A'B'} \equiv \frac{1}{c^2}T^{\mu\nu} + \Pi^{\mu\nu}.
\end{eqnarray}
\end{subequations}
We refer to such decomposition as the pre-ultralocal or pre-Carrollian parameterisation of $g_{\mu\nu}$, following \cite{Hansen:2021fxi,Banerjee:2025bkg}. In the second step, we define the Carroll expansion of pre-Carrollian variables,
\begin{subequations}
\begin{eqnarray}
	& T_{\mu}^A = \tau_{\mu}^A + c^2 \tau_{(2)\mu}^A + {\cal O}(c^4), \qquad & T^{\mu}_A = \tau^{\mu}_A + c^2 \tau^{(2)\mu}_A + {\cal O}(c^4), \\
	& E_{\mu}^{A'} = e_{\mu}^{A'} + c^2 e_{(2)\mu}^{A'} + {\cal O}(c^4), \qquad & E^{\mu}_A = e^{\mu}_{A'} + c^2 e^{(2)\mu}_{A'} + {\cal O}(c^4),
\end{eqnarray}
\end{subequations}
which gives the expansion of the pre-Carrollian variables,
\begin{subequations}\label{eq:pre-Carrollian T Pi c-sq exp}
\begin{eqnarray}
	& T_{\mu\nu} = \tau_{\mu\nu} + c^2 \tau_{(2)\mu\nu} + {\cal O}(c^4), \qquad & T^{\mu\nu} = \tau^{\mu\nu} + c^2 \tau_{(2)}^{\mu\nu} + {\cal O}(c^4), \\
	& \Pi_{\mu\nu} = h_{\mu\nu} + c^2 h_{(2)\mu\nu} + {\cal O}(c^4), \quad & \Pi^{\mu\nu} = h^{\mu\nu} + c^2 h_{(2)}^{\mu\nu} + {\cal O}(c^4),
\end{eqnarray}
\end{subequations}
where $\tau_{(2)\mu\nu} = (\tau_{\mu}^A\tau_{(2)\nu}^B + \tau_{(2)\mu}^A\tau_{\nu}^B)\eta_{AB}$ and so on. Then the Carroll expansion of the metric and its inverse is given by
\begin{equation}
	g_{\mu\nu} = \tau_{\mu\nu} + c^2 \big( h_{\mu\nu} + \tau_{(2)\mu\nu}\big) + {\cal O}(c^4), \quad g^{\mu\nu} = \frac{1}{c^2}\tau^{\mu\nu} + \big( h^{\mu\nu} + \tau_{(2)}^{\mu\nu}\big) + {\cal O}(c^2), \label{eq:string-Carroll expansion}
\end{equation}
which we refer to as the string-Carroll expansion of $g_{\mu\nu}$ and $g^{\mu\nu}$. The leading term gives the string-Carroll geometry described by $(h_{\mu\nu},\tau^{\mu\nu})$.

%%%%%%%%%%%%%%%%%%%%%
%%%% sec 3: magnetic string %%%%%

\section{Setting the stage:~The magnetic Carroll string}\label{sec:magnetic string}

In this section, we present an alternative description of the magnetic Carroll string introduced in \cite{Bagchi:2023cfp,Bagchi:2024rje}, inspired by \cite{Harksen:2024bnh}. To elaborate, in the Carroll expansion in \cite{Bagchi:2023cfp,Bagchi:2024rje}, the leading order (LO) and the next-to-leading order (NLO) Lagrangians describing the dynamics of the magnetic Carroll string are obtained either in the phase space form or in the Polyakov form, where the two are related by a Legendre transformation, \emph{i.e.}~integrating out the momentum from the phase space Lagrangian gives the Polyakov Lagrangian. In an alternate description, we start with the phase space Lagrangian for the relativistic string and integrate out only the longitudinal momentum while keeping the transverse momentum untouched and  express in it terms of a new field on the worldsheet. Then, Carroll expansion of this mixed form of the relativistic Lagrangian gives the LO Lagrangian that governs the magnetic Carroll string, whereas the NLO Lagrangian governs the dynamics of the sub-leading embedding fields in the magnetic Carroll string sector,
%\textcolor{red}{in a \emph{mixed phase space - Polyakov form}},
as we discuss below.

\subsection*{The pre-Carrollian magnetic string Lagrangian}

In the phase space formulation, the dynamics of a relativistic closed bosonic string moving in a Lorentzian target space with metric $g_{\mu\nu}$ is governed by the worldsheet action $S = \int d\sigma^0 d\sigma^1 {\cal L}$, where
\begin{equation}
	{\cal L} = P_{\mu}\dot{X}^{\mu} - \frac{e}{2}\big( c^2 g^{\mu\nu}P_{\mu}P_{\nu} + (c^2 T)^2 g_{\mu\nu}X'^{\mu}X'^{\nu}\big) - u P_{\mu}X'^{\mu} \label{eq:phase space L v1}
\end{equation}
is the phase space Lagrangian.\footnote{To be precise, $\oint d\sigma^1 {\cal L}$ is the Lagrangian. However, for the purposes of this paper, we refer to ${\cal L}$ as the Lagrangian.} $\sigma^{\hat{\alpha}} = (\sigma^0 = \tau, \sigma^1 = \sigma)$ are the worldsheet coordinates,
%which we take to be dimensionless.
and $\dot{X}$ and $X'$ denote derivatives of $X(\sigma^{\hat{\alpha}})$ with respect to $\tau$ and $\sigma$ respectively. $e(\sigma^{\hat{\alpha}})$ and $u(\sigma^{\hat{\alpha}})$ are Lagrange multipliers imposing the Hamiltonian constraints
\begin{equation}
	g^{\mu\nu}P_{\mu}P_{\nu} + (c T)^2 g_{\mu\nu}X'^{\mu}X'^{\nu} = 0, \qquad P_{\mu}X'^{\mu} = 0,
\end{equation}
respectively. The equation of motion for the momentum $P_{\mu}$ gives
\begin{equation}
	P_{\mu} = \frac{g_{\mu\nu}}{e c^2}\big(\dot{X}^{\nu} - u X'^{\nu}\big). \label{eq:rel momentum v1}
\end{equation}
%Using this equation of motion in the phase-space action gives the Polyakov Lagrangian
%\begin{equation}
%	{\cal L}_P = -\frac{T}{2}\sqrt{-\gamma}\gamma^{\alpha\beta}\partial_{\alpha}X^{\mu}\partial_{\beta}X^{\nu}g_{\mu\nu}(X).
%\end{equation}
Using the pre-Carrollian parameterisation of the metric, $g_{\mu\nu} = c^2 T_{\mu\nu} + \Pi_{\mu\nu}$, we can decompose the momentum into
%As discussed in section \ref{sec:string-Carroll expansion}, we can decompose the target space metric into the longitudinal metric $T_{\mu\nu}$ and the transverse metric $\Pi_{\mu\nu}$ as $g_{\mu\nu} = c^2 T_{\mu\nu} + \Pi_{\mu\nu}$. Substituting this in the expression for $P_{\mu}$, the momentum decomposes as
$P_{\mu} = P_{\mu}^{\parallel} + P_{\mu}^{\perp}$, where
\begin{equation}
	P_{\mu}^{||} \equiv P_{\nu}T^{\nu}_{\mu} = \frac{T_{\mu\nu}}{e}\big(\dot{X}^{\nu} - u X'^{\nu}\big), \qquad P_{\mu}^{\perp} \equiv P_{\nu}\Pi^{\nu}_{\mu} = \frac{\Pi_{\mu\nu}}{e c^2}\big(\dot{X}^{\nu} - u X'^{\nu}\big), \label{eq:long-trans momenta v1}
\end{equation}
are the longitudinal and transverse components of the momentum that satisfy $P_{\mu}^{||}\Pi^{\mu\nu} = 0$ and $P_{\mu}^{\perp}T^{\mu\nu} = 0$. Here, $T^{\mu}_{\nu} \equiv T^{\mu\lambda}T_{\lambda\nu}$ and $\Pi^{\mu}_{\nu} \equiv \Pi^{\mu\lambda}\Pi_{\lambda\nu}$ are pre-Carrollian longitudinal and transverse projectors respectively. Substituting the decomposition of the momentum in the phase space Lagrangian \eqref{eq:phase space L v1}, it also decomposes into a longitudinal part ${\cal L}^{\parallel}$ and a transverse part ${\cal L}^{\perp}$, \emph{i.e.}~${\cal L} = {\cal L}^{\parallel}  + {\cal L}^{\perp}$, where
\begin{eqnarray}
	{\cal L}^{\parallel}  &=& P_{\mu}^{\parallel}\big(\dot{X}^{\mu} - u X'^{\mu}\big) - \frac{e}{2}\big(T^{\mu\nu}P_{\mu}^{\parallel}P_{\nu}^{\parallel} + c^6 T^2 T_{\mu\nu}X'^{\mu}X'^{\nu}\big), \label{eq:long phase space L v1} \\
	{\cal L}^{\perp}  &=& P_{\mu}^{\perp}\big(\dot{X}^{\mu} - u X'^{\mu}\big) - \frac{e}{2}\big(c^2 \Pi^{\mu\nu}P_{\mu}^{\perp}P_{\nu}^{\perp} + c^4 T^2 \Pi_{\mu\nu}X'^{\mu}X'^{\nu}\big). \label{eq:trans phase space L v1}
\end{eqnarray}
%In the formulation of magnetic Carroll string in \cite{Bagchi:2024rje}, we Carroll expand $P_{\mu}^{||}$, $P_{\mu}^{\perp}$ and integrate out both $P_{\mu}^{||}$, $P_{\mu}^{\perp}$ at all orders in $c^2$ to get the leading order and next-to-leading order Polyakov Lagrangians for the magnetic Carroll string. Here, instead, we only integrate out the longitudinal momentum and not the transverse momentum.
We first integrate out the longitudinal momentum $P_{\mu}^{\parallel}$ from the longitudinal Lagrangian ${\cal L}^{\parallel}$ using \eqref{eq:long-trans momenta v1}, which gives
\begin{equation}
	{\cal L}^{\parallel} = \frac{1}{2e}\big(\dot{X}^{\mu}\dot{X}^{\nu} - 2u\dot{X}^{\mu}X'^{\nu} + (u^2 - e^2(c^3T)^2)X'^{\mu}X'^{\nu}\big)T_{\mu\nu}. \label{eq:long config L}
\end{equation}
Rescaling the tension $T_{\eff} = c^3 T$ and the Lagrange multiplier $e_{\eff} = e$ such that $T_{\eff}$, $e_{\eff}$ remain finite as $c\rightarrow 0$, we can define a Lorentzian fiducial metric on the worldsheet,
\begin{equation}
	\gamma^{\hat{\alpha}\hat{\beta}} = \begin{pmatrix}
		-1 & u \\ u & -u^2 + e_{\eff}^2 T_{\eff}^2
	\end{pmatrix}, \qquad \sqrt{-\gamma} = \frac{1}{e_{\eff} T_{\eff}}, \label{eq:rel fiducial metric v1}
\end{equation}
which enables us to write the longitudinal Lagrangian \eqref{eq:long config L} in the Polyakov form,
\begin{equation}
	{\cal L}^{\parallel} = -\frac{T_{\eff}}{2}\sqrt{-\gamma}\gamma^{\hat{\alpha}\hat{\beta}}\partial_{\hat{\alpha}}X^{\mu}\partial_{\hat{\beta}}X^{\nu}T_{\mu\nu}. \label{eq:long Polyakov L v1}
\end{equation}
The fiducial metric \eqref{eq:rel fiducial metric v1} can also be used to write the transverse momentum in a covariant form,
\begin{equation}
	P_{\mu}^{\perp} = -\frac{\gamma^{\tau\hat{\alpha}}\partial_{\hat{\alpha}}X^{\nu}\Pi_{\nu\mu}}{e c^2}. \label{eq:trans P v1}
\end{equation}
As mentioned earlier, we do not integrate out the transverse momentum $P_{\mu}^{\perp}$ from the transverse Lagrangian ${\cal L}^{\perp}$. Instead, we rewrite $P_{\mu}^{\perp}$ in terms of a field $\kappa_{\hat{\alpha}}^{\mu}(\tau,\sigma)$ on the worldsheet,
\begin{equation}
	P_{\mu}^{\perp} = -\frac{\gamma^{\tau\hat{\alpha}}\kappa_{\hat{\alpha}}^{\nu}\Pi_{\nu\mu}}{e}; \qquad \kappa_{\hat{\alpha}}^{\mu} \equiv \frac{\partial_{\hat{\alpha}}X^{\nu}\Pi^{\mu}_{\nu}}{c^2}. \label{eq:trans P HHST}
\end{equation}
Substituting the expression \eqref{eq:trans P HHST} for $P_{\mu}^{\perp}$ in the transverse Lagrangian \eqref{eq:trans phase space L v1} and also using the fiducial metric \eqref{eq:rel fiducial metric v1}, we can write ${\cal L}^{\perp}$ in a covariant form,
\begin{equation}
	{\cal L}^{\perp} = -\frac{T_{\eff}}{2}\sqrt{-\gamma}\gamma^{\hat{\alpha}\hat{\beta}}\big(2\kappa_{\hat{\alpha}}^{\mu}\partial_{\hat{\beta}}X^{\nu}\Pi_{\mu\nu} - c^2 \kappa_{\hat{\alpha}}^{\mu}\kappa_{\hat{\beta}}^{\nu}\Pi_{\mu\nu}\big). \label{eq:transv Polyakov L v1}
\end{equation}
%\begin{eqnarray}
%	{\cal L}^{\perp} &=& P_{\mu}^{\perp}\big(\dot{X}^{\mu} - u X'^{\mu}\big) - \frac{e}{2}\big(c^2 \Pi^{\mu\nu}P_{\mu}^{\perp}P_{\nu}^{\perp} + c^4 T^2 \Pi_{\mu\nu}X'^{\mu}X'^{\nu}\big) \nonumber \\
%	&=& -\frac{\Pi_{\mu\nu}}{e}(-\kappa_{\tau}^{\mu} + u\kappa_{\sigma}^{\mu})(\dot{X}^{\nu} - u X'^{\nu}) - \frac{c^2\Pi_{\mu\nu}}{2e}(-\kappa_{\tau}^{\mu} + u\kappa_{\sigma}^{\mu})(-\kappa_{\tau}^{\nu} + u\kappa_{\sigma}^{\nu}) \nonumber \\
%	&& - \frac{e c^4 T^2}{2}\Pi_{\mu\nu}X'^{\mu}X'^{\nu} \nonumber \\
%	&=& -\frac{\Pi_{\mu\nu}}{e}\gamma^{\alpha\beta}\kappa_{\alpha}^{\mu}\partial_{\beta}X^{\nu} + \frac{c^2\Pi_{\mu\nu}}{2e}\gamma^{\alpha\beta}\kappa_{\alpha}^{\mu}\kappa_{\beta}^{\nu} + \Big(\Pi_{\mu\nu} e c^6 T^2 \kappa_{\sigma}^{\mu}X'^{\nu} - \frac{c^2}{2}\Pi_{\mu\nu} e c^6 T^2\kappa_{\sigma}^{\mu}\kappa_{\sigma}^{\nu} \nonumber \\
%	&& - \frac{e c^4 T^2}{2}\Pi_{\mu\nu}X'^{\mu}X'^{\nu}\Big) \nonumber \\
%	&=& -\frac{T_{\eff}}{2}\sqrt{-\gamma}\gamma^{\alpha\beta}\big(2\kappa_{\alpha}^{\mu}\partial_{\beta}X^{\nu}\Pi_{\mu\nu} - c^2 \kappa_{\alpha}^{\mu}\kappa_{\beta}^{\nu}\Pi_{\mu\nu}\big).
%\end{eqnarray}
Finally, putting the longitudinal Lagrangian \eqref{eq:long Polyakov L v1} and the transverse Lagrangian \eqref{eq:transv Polyakov L v1} together, we get the full Lagrangian
\begin{equation}
	{\cal L}^{(m)} = -\frac{T_{\eff}}{2}\sqrt{-\gamma}\gamma^{\hat{\alpha}\hat{\beta}}\big(\partial_{\hat{\alpha}}X^{\mu}\partial_{\hat{\beta}}X^{\nu}T_{\mu\nu} + 2\kappa_{\hat{\alpha}}^{\mu}\partial_{\hat{\beta}}X^{\nu}\Pi_{\mu\nu} - c^2 \kappa_{\hat{\alpha}}^{\mu}\kappa_{\hat{\beta}}^{\nu}\Pi_{\mu\nu}\big), \label{eq:pre-Carrollian mag L v1}
\end{equation}
which we refer to as the \emph{pre-Carrollian magnetic string Lagrangian}.

%\textcolor{red}{We would like to emphasise that though the transverse Lagrangian \eqref{eq:transv Polyakov L v1} may appear to be in a Polyakov-like form, the field $\kappa_{\hat{\alpha}}^{\mu}$ is essentially the (unintegrated) transverse momentum $P_{\mu}^{\perp}$. Thus, we refer to the Lagrangian \eqref{eq:pre-Carrollian mag L v1} to be in a \emph{mixed phase space - Polyakov form}, to distinguish it from the true Polyakov Lagrangian.}
%in \cite{Bagchi:2024rje}.
%This is the curved target space generalisation of the pre-Carrollian magnetic Lagrangian found in \cite{Harksen:2024bnh}.
%\subsubsection{Consistency check}
%
%The equation of motion for the field $\kappa_{\alpha}^{\mu}$ gives $\kappa_{\beta}^{\nu}\Pi_{\nu\mu} = \frac{1}{c^2}\partial_{\beta}X^{\nu}\Pi_{\nu\mu}$, which is consistent with our previous definition \eqref{}. Integrating out $\kappa_{\alpha}^{\mu}$ from ${\cal L}^{(m)}$, it reduces to
%\begin{eqnarray}
%	{\cal L} &=& -\frac{c^3 T}{2}\sqrt{-\gamma}\gamma^{\alpha\beta}\big(\partial_{\alpha}X^{\mu}\partial_{\beta}X^{\nu}T_{\mu\nu} + \frac{1}{c^2}\partial_{\alpha}X^{\mu}\partial_{\beta}X^{\nu}\Pi_{\mu\nu}\big) \nonumber \\
%	&=& -\frac{cT}{2}\sqrt{-\gamma}\gamma^{\alpha\beta}\big(\partial_{\alpha}X^{\mu}\partial_{\beta}X^{\nu}T_{\mu\nu}c^2 + \partial_{\alpha}X^{\mu}\partial_{\beta}X^{\nu}\Pi_{\mu\nu}\big) \nonumber \\
%	&=& -\frac{cT}{2}\sqrt{-\gamma}\gamma^{\alpha\beta}\partial_{\alpha}X^{\mu}\partial_{\beta}X^{\nu}g_{\mu\nu} \nonumber \\
%	&=& {\cal L}_P,
%\end{eqnarray}
%which is the Polyakov Lagrangian for the relativistic closed bosonic tensile string.

\subsection*{Carroll expansion: LO and NLO Lagrangians}

The Carroll expansion of the embedding fields $X^{\mu}(\sigma^{\hat{\alpha}})$ in powers of $c^2$,
\begin{equation}
	X^{\mu}(\sigma^{\hat{\alpha}}) = x^{\mu}(\sigma^{\hat{\alpha}}) + c^2 y^{\mu}(\sigma^{\hat{\alpha}}) + c^4 z^{\mu}(\sigma^{\hat{\alpha}}) + \calO(c^6) \label{eq:c-sq exp X}
\end{equation}
induces a Taylor expansion on the $c^2$ expansion \eqref{eq:pre-Carrollian T Pi c-sq exp} of $T_{\mu\nu}(X)$ and $\Pi_{\mu\nu}(X)$ giving
\begin{subequations}
	\begin{eqnarray}
		T_{\mu\nu}(X) &=& \tau_{\mu\nu}(x) + c^2 \tau_{(2)\mu\nu}(x,y) + \calO(c^4), \\
		\Pi_{\mu\nu}(X) &=& h_{\mu\nu}(x) + c^2 h_{(2)\mu\nu}(x,y) + \calO(c^4),
	\end{eqnarray} \label{eq:c-sq exp T Pi}
\end{subequations}
where
\begin{equation}
	\tau_{(2)\mu\nu}(x,y) \equiv \tau_{(2)\mu\nu}(x) + y^{\rho}\partial_{\rho}\tau_{\mu\nu}(x), \qquad h_{(2)\mu\nu}(x,y) \equiv h_{(2)\mu\nu}(x) + y^{\rho}\partial_{\rho}h_{\mu\nu}(x).
\end{equation}
The Carroll expansion of the worldsheet fiducial metric\footnote{The Lagrange multipliers $e_{\eff}$ and $u$ are Carroll expanded as $e_{\eff} = e_{\eff (0)} + {\cal O}(c^2)$ and $u = u_{(0)} + {\cal O}(c^2)$ giving
\begin{equation}\label{eq:rel LO fiducial metric v1}
	\gamma_{(0)}^{\hat{\alpha}\hat{\beta}} =
	\begin{pmatrix}
		-1 & u_{(0)} \\ u_{(0)} & -u_{(0)}^2 + e_{\eff(0)}^2 T_{\eff}^2
	\end{pmatrix}.
\end{equation}} and the field $\kappa_{\hat{\alpha}}^{\mu}$ is given by
\begin{equation}
	\kappa_{\hat{\alpha}}^{\mu} = \kappa_{(0)\hat{\alpha}}^{\mu} + c^2 \kappa_{(0)\hat{\alpha}}^{\mu} + \calO(c^4), \qquad \gamma_{\hat{\alpha}\hat{\beta}} = \gamma_{(0)\hat{\alpha}\hat{\beta}} + c^2 \gamma_{(2)\hat{\alpha}\hat{\beta}} + \calO(c^4), \label{eq:c-sq exp gamma}
\end{equation}
%\begin{subequations}
%	\begin{eqnarray}
%	\kappa_{\hat{\alpha}}^{\mu} &=&  \kappa_{(0)\hat{\alpha}}^{\mu} + c^2 \kappa_{(0)\hat{\alpha}}^{\mu} + \calO(c^4), \qquad \gamma_{\hat{\alpha}\hat{\beta}} = \gamma_{(0)\hat{\alpha}\hat{\beta}} + c^2 \gamma_{(2)\hat{\alpha}\hat{\beta}} + \calO(c^4), \\
%	\gamma^{\hat{\alpha}\hat{\beta}} &=& \gamma_{(0)}^{\hat{\alpha}\hat{\beta}} - c^2 \gamma_{(2)}^{\hat{\alpha}\hat{\beta}} + \calO(c^4), \qquad \sqrt{-\gamma} = \sqrt{-\gamma_{(0)}}\Big(1 + \frac{c^2}{2}\gamma_{(2)\hat{\alpha}}^{\hat{\alpha}}\Big) + \calO(c^4),
%\end{eqnarray}
%\end{subequations}
and $\gamma^{\hat{\alpha}\hat{\beta}} = \gamma_{(0)}^{\hat{\alpha}\hat{\beta}} - c^2 \gamma_{(2)}^{\hat{\alpha}\hat{\beta}} + \calO(c^4)$ with $\gamma_{(0)\hat{\alpha}\hat{\beta}}\gamma_{(0)}^{\hat{\beta}\hat{\rho}} = \delta^{\hat{\rho}}_{\hat{\alpha}}$ and $\gamma_{(2)}^{\hat{\alpha}\hat{\beta}} = \gamma_{(0)}^{\hat{\alpha}\hat{\rho}}\gamma_{(0)}^{\hat{\beta}\hat{\lambda}}\gamma_{(2)\hat{\rho}\hat{\lambda}}$.
%and $\gamma_{(2)\hat{\alpha}}^{\hat{\alpha}} = \gamma_{(0)}^{\hat{\alpha}\hat{\beta}}\gamma_{(2)\hat{\alpha}\hat{\beta}}$.
Using these Carroll expansions in \eqref{eq:pre-Carrollian mag L v1}, the pre-Carrollian magnetic string Lagrangian expands in powers of $c^2$ as
\begin{equation}
	{\cal L}^{(m)} = {\cal L}^{(m)}_{LO} + c^2 {\cal L}^{(m)}_{NLO} + \calO(c^4),
\end{equation}
where the leading order (LO) Lagrangian is given by
\begin{equation}
	{\cal L}^{(m)}_{LO} =  -\frac{T_{\eff}}{2}\sqrt{-\gamma_{(0)}}\gamma_{(0)}^{\hat{\alpha}\hat{\beta}}\big(\partial_{\hat{\alpha}}x^{\mu}\partial_{\hat{\beta}}x^{\nu}\tau_{\mu\nu} + 2\ost{\kappa}{\zero}_{\hat{\alpha}}^{\mu}\partial_{\hat{\beta}}x^{\nu}h_{\mu\nu}\big). \label{eq:mag LO L v1}
\end{equation}
The LO Lagrangian describes the magnetic Carroll string moving in a curved string-Carroll target spacetime. The flat string-Carroll target spacetime form of this Lagrangian was obtained from an equivalent Carroll limit in \cite{Harksen:2024bnh}.

The next-to-leading order (NLO) Lagrangian that governs the dynamics of sub-leading embedding fields $y^{\mu}(\sigma^{\hat{\alpha}})$ is given by
\begin{eqnarray}
	{\cal L}^{(m)}_{NLO} &=& -\frac{T_{\eff}}{2}\sqrt{-\gamma_{(0)}}\Big[\gamma_{(0)}^{\hat{\alpha}\hat{\beta}}\big(2\partial_{\hat{\alpha}}x^{\mu}\partial_{\hat{\beta}}y^{\nu}\tau_{\mu\nu} + \partial_{\hat{\alpha}}x^{\mu}\partial_{\hat{\beta}}x^{\nu}\tau_{(2)\mu\nu}(x,y) \nonumber \\
	&& + 2\kappa_{(2)\hat{\alpha}}^{\mu}\partial_{\hat{\beta}}x^{\nu}h_{\mu\nu} + 2\kappa_{(0)\hat{\alpha}}^{\mu}\partial_{\hat{\beta}}y^{\nu}h_{\mu\nu} + 2\kappa_{(0)\hat{\alpha}}^{\mu}\partial_{\hat{\beta}}x^{\nu}h_{(2)\mu\nu}(x,y) - \kappa_{(0)\hat{\alpha}}^{\mu}\kappa_{(0)\hat{\beta}}^{\nu}h_{\mu\nu}\big) \nonumber \\
	&& - \half G_{(0)}^{\hat{\alpha}\hat{\beta}\rho\lambda}\gamma_{(2)\rho\lambda}\big(\partial_{\hat{\alpha}}x^{\mu}\partial_{\hat{\beta}}x^{\nu}\tau_{\mu\nu} + 2\kappa_{(0)\hat{\alpha}}^{\mu}\partial_{\hat{\beta}}x^{\nu}h_{\mu\nu}\big)\Big], \label{eq:mag L NLO v1}
\end{eqnarray}
where we have used the Wheeler de Witt metric
\begin{equation}
	G_{(0)}^{\hat{\alpha}\hat{\beta}\hat{\rho}\hat{\lambda}} = \gamma_{(0)}^{\hat{\alpha}\hat{\rho}}\gamma_{(0)}^{\hat{\beta}\hat{\lambda}} + \gamma_{(0)}^{\hat{\alpha}\hat{\lambda}}\gamma_{(0)}^{\hat{\beta}\hat{\rho}} - \gamma_{(0)}^{\hat{\alpha}\hat{\beta}}\gamma_{(0)}^{\hat{\rho}\hat{\lambda}}.
\end{equation}

\subsection*{Comparing the two descriptions}

In the two descriptions for the magnetic Carroll string, \ie~\eqref{eq:mag LO L v1} and that in \cite{Bagchi:2024rje}, we rescale the tension $T$ and the Lagrange multiplier $e$,
\begin{subequations}
	\begin{eqnarray}
	e_{\eff} = e, \quad T_{\eff} = c^3T, \quad &\implies& \quad e_{\eff}T_{\eff} = c^6 e^2 T^2, \\
	\tilde{e} = c^2 e, \quad \tilde{T} = cT, \quad &\implies& \quad \tilde{e}\tilde{T} = c^6 e^2 T^2,
\end{eqnarray}
\end{subequations}
such that the product $\tilde{e}\tilde{T} = e_{\eff}T_{\eff}$ remains fixed in the $c\rightarrow 0$ limit giving the Lorentzian fiducial metric \eqref{eq:rel LO fiducial metric v1} on the worldsheet.
%Both descriptions give the same dynamics of the magnetic Carroll string, and
In appendix~\ref{app:magnetic string near black hole}, we verify that both the descriptions are equivalent and give the same dynamics of the magnetic Carroll string by comparing the equations of motion for the LO and NLO embedding fields $x^{\mu}$ and $y^{\mu}$, respectively,  for the magnetic Carroll string propagating in the near-horizon region of the Schwarzschild black hole. The difference arises in offshell actions:~in \cite{Bagchi:2024rje}, the LO transverse fields $x^i$ for $i=2,\dots,D-1$ appear in the LO action and the LO longitudinal fields $x^a$ for $a=0,1$ appear in the NLO action, whereas both longitudinal and transverse fields appear in the LO action \eqref{eq:mag LO L v1}; with the same pattern repeating at higher orders.

%%%%%%%%%%%%%%%%%%%%%%%%%%%%%
%%%% Sec-3: Chiral expansion %%%%%%%%%%%%

\section{Chiral Carroll string from small-$c$ expansion}\label{sec:chiral string from expansion}

The description of the magnetic Carroll string in the previous section provides a setup to define the chiral Carroll string. As before, we do not integrate out the transverse momentum $P_{\mu}^{\perp}$ from the phase space Lagrangian but now decompose it into two fields of opposite chirality. Inspired by \cite{Harksen:2024bnh}, we do so by keeping the two fields at different orders in $c^2$, thus introducing chirality in the theory when we do the Carroll expansion.

\subsection*{The pre-Carrollian chiral string Lagrangian}

We define chiral projectors on the worldsheet of a relativistic string,
$\Gamma^{\hat{\alpha}\hat{\beta}}_{\pm} \equiv \dfrac{1}{2}(\gamma^{\hat{\alpha}\hat{\beta}} \pm \varepsilon^{\hat{\alpha}\hat{\beta}})$, where the antisymmetric symbol $\varepsilon^{\hat{\alpha}\hat{\beta}}$ is normalised with respect to $\boldsymbol{\varepsilon}^{\hat{\alpha}\hat{\beta}}$ on a flat worldsheet as $\varepsilon^{\hat{\alpha}\hat{\beta}} \equiv \dfrac{\boldsymbol{\varepsilon}^{\hat{\alpha}\hat{\beta}}}{\sqrt{-\gamma}}$ with $\boldsymbol{\varepsilon}^{\tau\sigma} = 1$. Using these chiral projectors, we introduce a chiral field $\psi_{\hat{\alpha}}^{\mu}(\tau,\sigma)$ and an anti-chiral field $\chi_{\hat{\alpha}}^{\mu}(\tau,\sigma)$ on the worldsheet,
\begin{equation}
	\psi_{\hat{\alpha}}^{\mu} \equiv \frac{\gamma_{\hat{\alpha}\hat{\rho}}\Gamma^{\hat{\rho}\hat{\lambda}}_{-}\partial_{\hat{\lambda}}X^{\nu}\Pi^{\mu}_{\nu}}{c^2}, \quad \chi_{\hat{\alpha}}^{\mu} \equiv \frac{\gamma_{\hat{\alpha}\hat{\rho}}\Gamma^{\hat{\rho}\hat{\lambda}}_{+}\partial_{\hat{\lambda}}X^{\nu}\Pi^{\mu}_{\nu}}{c^4}.
\end{equation}
These fields satisfy
\begin{equation}\label{eq:psi-psi-equivalence}
	\psi_{\hat{\alpha}}^{\mu} = \gamma_{\hat{\alpha}\hat{\rho}}\Gamma^{\hat{\rho}\hat{\lambda}}_{-}\psi_{\hat{\lambda}}^{\mu}, \quad \chi_{\hat{\alpha}}^{\mu} = \gamma_{\hat{\alpha}\hat{\rho}}\Gamma_{+}^{\hat{\rho}\hat{\lambda}}\chi_{\hat{\lambda}}^{\mu},
\end{equation}
which follows from the property,\footnote{Another identity that has been used in some calculations is $\varepsilon^{\hat{\alpha}\hat{\rho}}\gamma_{\hat{\rho}\hat{\lambda}}\varepsilon^{\hat{\lambda}\hat{\beta}} = \gamma^{\hat{\alpha}\hat{\beta}}$.}
\begin{equation}
	(\Gamma_{\pm}^2)^{\hat{\alpha}\hat{\beta}} = \Gamma^{\hat{\alpha}\hat{\rho}}_{\pm}\gamma_{\hat{\rho}\hat{\lambda}}\Gamma^{\hat{\lambda}\hat{\beta}}_{\pm} = \Gamma^{\hat{\alpha}\hat{\beta}}_{\pm}.
\end{equation}
Then, we can decompose the transverse momentum \eqref{eq:trans P v1} into chiral and anti-chiral parts, and express it in terms of $\psi_{\hat{\alpha}}^{\mu}$, $\chi_{\hat{\alpha}}^{\mu}$ as
%\begin{equation}
%	P_{\mu}^{\perp} = - \frac{1}{e}\big(\Gamma^{\tau\hat{\alpha}}_{-} + \Gamma^{\tau\hat{\alpha}}_{+}\big)\frac{\partial_{\hat{\alpha}}X^{\nu}\Pi_{\nu\mu}}{c^2}. \label{eq:trans P split v1}
%\end{equation}
\begin{equation}
	P_{\mu}^{\perp} = -\frac{\Pi_{\mu\nu}}{e}\gamma^{\tau\hat{\alpha}}\big(\psi_{\hat{\alpha}}^{\nu} + c^2 \chi_{\hat{\alpha}}^{\nu}\big). \label{eq:trans P chiral split v1}
\end{equation}
Substituting $P_{\mu}^{\perp}$ in the transverse Lagrangian \eqref{eq:trans phase space L v1}, we get
\begin{equation}
	{\cal L}^{\perp} = -T_{\eff}\sqrt{-\gamma}\big(\Gamma_{+}^{\hat{\alpha}\hat{\beta}}\psi_{\hat{\alpha}}^{\mu}\partial_{\hat{\beta}}X^{\nu}\Pi_{\mu\nu} + c^2 \Gamma_{-}^{\hat{\alpha}\hat{\beta}}\chi_{\hat{\alpha}}^{\mu}\partial_{\hat{\beta}}X^{\nu}\Pi_{\mu\nu} - c^4 \gamma^{\hat{\alpha}\hat{\beta}}\psi_{\hat{\alpha}}^{\mu}\chi_{\hat{\beta}}^{\nu}\Pi_{\mu\nu}\big),
\end{equation}
where $T_{\eff} = c^3 T$ as defined for the magnetic string.
The longitudinal sector is identical to that of the magnetic string, \ie~\eqref{eq:long Polyakov L v1}, which together with the transverse Lagrangian gives
\begin{eqnarray}
	{\cal L}^{(c)} &=& -\frac{T_{\eff}}{2}\sqrt{-\gamma}\big(\gamma^{\hat{\alpha}\hat{\beta}}\partial_{\hat{\alpha}}X^{\mu}\partial_{\hat{\beta}}X^{\nu}T_{\mu\nu} + 2\Gamma_{+}^{\hat{\alpha}\hat{\beta}}\psi_{\hat{\alpha}}^{\mu}\partial_{\hat{\beta}}X^{\nu}\Pi_{\mu\nu} \nonumber \\
	&& + 2c^2 \Gamma_{-}^{\hat{\alpha}\hat{\beta}}\chi_{\hat{\alpha}}^{\mu}\partial_{\hat{\beta}}X^{\nu}\Pi_{\mu\nu} - 2c^4 \gamma^{\hat{\alpha}\hat{\beta}}\psi_{\hat{\alpha}}^{\mu}\chi_{\hat{\beta}}^{\nu}\Pi_{\mu\nu}\big). \label{eq:pre-Carrollian chiral L v1}
\end{eqnarray}
We refer to ${\cal L}^{(c)}$ as the \emph{pre-Carrollian chiral string Lagrangian}, where the $\chi_{\hat{\alpha}}^{\mu}$-dependent term appears at one higher order in $c^2$ than the $\psi_{\hat{\alpha}}^{\mu}$-dependent term, as desired.
%As a consistency check, we can check that the equations of motion for $\psi_{\hat{\alpha}}^{\mu}$ and $\chi_{\hat{\alpha}}^{\nu}$ give \eqref{}, which upon substituting in ${\cal L}^{(c)}$ gives back the Polyakov Lagrangian ${\cal L}_P$.

\subsection*{Carroll expansion}
%\subsection{Carroll expansion of pre-Carrollian chiral string Lagrangian}

 The Carroll expansion of $X^{\mu}(\tau,\sigma)$, $T_{\mu\nu}(X)$, $\Pi_{\mu\nu}(X)$, and $\gamma_{\hat{\alpha}\hat{\beta}}(\tau,\sigma)$ is given in \eqref{eq:c-sq exp X}-\eqref{eq:c-sq exp gamma}, and the Carroll expansion of $\psi_{\hat{\alpha}}^{\mu}(\tau,\sigma)$ and $\chi_{\hat{\alpha}}^{\mu}(\tau,\sigma)$ is given by
\begin{equation}
	\psi_{\hat{\alpha}}^{\mu} = \psi_{(0)\hat{\alpha}}^{\mu} + c^2 \psi_{(2)\hat{\alpha}}^{\mu} + \calO(c^4), \quad \chi_{\hat{\alpha}}^{\mu} = \chi_{(0)\hat{\alpha}}^{\mu} + c^2 \chi_{(2)\hat{\alpha}}^{\mu} + \calO(c^4).
\end{equation}
Due to the Carroll expansion of $\gamma_{\hat{\alpha}\hat{\beta}}$, the antisymmetric symbol expands as
\begin{eqnarray}
	\varepsilon^{\hat{\alpha}\hat{\beta}} = \frac{\boldsymbol{\varepsilon}^{\hat{\alpha}\hat{\beta}}}{\sqrt{-\gamma_{(0)}}} - c^2\Big(\frac{\boldsymbol{\varepsilon}^{\hat{\alpha}\hat{\beta}}}{\sqrt{-\gamma_{(0)}}}\frac{\gamma_{(0)}^{\hat{\rho}\hat{\lambda}}\gamma_{(2)\hat{\rho}\hat{\lambda}}}{2}\Big) + \calO(c^4) \equiv \varepsilon_{(0)}^{\hat{\alpha}\hat{\beta}} - c^2 \varepsilon_{(2)}^{\hat{\alpha}\hat{\beta}} + \calO(c^4),
\end{eqnarray}
which gives the Carroll expansion of the chiral projectors,
\begin{eqnarray}
	\Gamma_{\pm}^{\hat{\alpha}\hat{\beta}} = \frac{1}{2}\big(\gamma_{(0)}^{\hat{\alpha}\hat{\beta}} \pm \varepsilon_{(0)}^{\hat{\alpha}\hat{\beta}}\big) - c^2 \frac{1}{2}\big(\gamma_{(0)}^{\hat{\alpha}\hat{\beta}} \pm \varepsilon_{(2)}^{\hat{\alpha}\hat{\beta}}\big) + \calO(c^4) \equiv \Gamma_{(0)\pm}^{\hat{\alpha}\hat{\beta}} - c^2 \Gamma_{(2)\pm}^{\hat{\alpha}\hat{\beta}} + \calO(c^4).
\end{eqnarray}
Then using all these Carroll expansions in the pre-Carrollian chiral string Lagrangian \eqref{eq:pre-Carrollian chiral L v1}, it expands as
\begin{equation}
	{\cal L}^{(c)} = {\cal L}^{(c)}_{LO} + c^2 {\cal L}^{(c)}_{NLO} + \calO(c^4),
\end{equation}
where the leading order (LO) Lagrangian is given by
\begin{equation}
	{\cal L}^{(c)}_{LO} =  -\frac{T_{\eff}}{2}\sqrt{-\gamma_{(0)}}\big(\gamma_{(0)}^{\hat{\alpha}\hat{\beta}}\partial_{\hat{\alpha}}x^{\mu}\partial_{\hat{\beta}}x^{\nu}\tau_{\mu\nu} + 2\Gamma_{(0)+}^{\hat{\alpha}\hat{\beta}}\psi_{(0)\hat{\alpha}}^{\mu}\partial_{\hat{\beta}}x^{\nu}h_{\mu\nu}\big). \label{eq:chiral L LO}
\end{equation}
It only contains the (LO) chiral field $\psi_{(0)\hat{\alpha}}^{\mu}$, which forces the transverse embedding fields $x^i$ to be functions of either the right-moving or the left-moving coordinate on the worldsheet, as we show in detail in section~\ref{sec:chiral string near black hole}. Thus, we refer to the LO Lagrangian ${\cal L}^{(c)}_{LO}$ as describing the \emph{chiral Carroll string}. Its specialisation to the flat string-Carroll target spacetime was obtained from an equivalent Carroll limit in \cite{Harksen:2024bnh}.

The next-to-leading order Lagrangian is given by
\begin{eqnarray}
	\hspace{-1.5mm} {\cal L}^{(c)}_{NLO} &=& -\frac{T_{\eff}}{2}\sqrt{-\gamma_{(0)}}\Big[\gamma_{(0)}^{\hat{\alpha}\hat{\beta}}\big(2\partial_{\hat{\alpha}}x^{\mu}\partial_{\hat{\beta}}y^{\nu}\tau_{\mu\nu} + \partial_{\hat{\alpha}}x^{\mu}\partial_{\hat{\beta}}x^{\nu}\tau_{(2)\mu\nu}(x,y)\big) \nonumber \\
	&& + 2\Gamma_{(0)+}^{\hat{\alpha}\hat{\beta}}\big(\psi_{(2)\hat{\alpha}}^{\mu}\partial_{\hat{\beta}}x^{\nu}h_{\mu\nu} + \psi_{(0)\hat{\alpha}}^{\mu}\partial_{\hat{\beta}}y^{\nu}h_{\mu\nu} + \psi_{(0)\hat{\alpha}}^{\mu}\partial_{\hat{\beta}}x^{\nu}h_{(2)\mu\nu}(x,y)\big) \nonumber \\
	&& + 2\Gamma_{(0)-}^{\hat{\alpha}\hat{\beta}} \chi_{(0)\hat{\alpha}}^{\mu}\partial_{\hat{\beta}}x^{\nu}h_{\mu\nu} - \half G_{(0)}^{\hat{\alpha}\hat{\beta}\hat{\rho}\hat{\lambda}}\gamma_{(2)\hat{\rho}\hat{\lambda}}\big(\partial_{\hat{\alpha}}x^{\mu}\partial_{\hat{\beta}}x^{\nu}\tau_{\mu\nu} + \psi_{(0)\hat{\alpha}}^{\mu}\partial_{\hat{\beta}}x^{\nu}h_{\mu\nu}\big)\Big], \label{eq:chiral L NLO}
\end{eqnarray}
which also contains the (LO) anti-chiral field $\chi_{(0)\hat{\alpha}}^{\mu}$.

%\subsection*{Symmetries}
%
%The worldsheet has diffeomorphisms or reparameterisations, and Weyl rescaling symmetry.
%
%The target space has string-Carroll symmetries. \textcolor{red}{talk about chiral enhancement of symmetries}

%%%%%%%%%%%%%%%%%%%%%%%%%%%%%
%%% Sec-4: Chiral string near black holes %%%%%%%

\section{Chiral Carroll string near Schwarzschild horizon}\label{sec:chiral string near black hole}

In \cite{Bagchi:2023cfp,Bagchi:2024rje}, it was shown that the near-horizon geometry of the Schwarzschild black hole in $4$-dimensional asymptotically flat spacetime can be formulated as a string-Carroll expansion discussed in section~\ref{sec:string-Carroll expansion}, with the distance away from the horizon playing the role of the Carrollian expansion parameter $c^2$.\footnote{See also \cite{Grumiller:2019fmp,Fontanella:2022gyt,Bagchi:2026qpi} for discussions on near-horizon expansions of non-extremal black holes and black branes including the Kerr black hole, AdS black brane, Lifshitz black hole, and associated non-Lorentzian structures.} Then, in these papers, it was found that an infalling relativistic closed bosonic string behaves as an electric Carroll string or a magnetic Carroll string moving in the near-horizon string-Carroll expanded geometry as the Carrollian target spacetime. In this section, we analyse the dynamics of an infalling relativistic closed bosonic string when it behaves as a chiral Carroll string as it approaches the horizon. In particular, we use the setup of chiral Carroll strings developed in section~\ref{sec:chiral string from expansion} with the Carrollian expansion parameter $c^2$ replaced by a near-horizon distance parameter, and find the equations of motion and their solutions for the LO embedding fields.

\subsection{Near-horizon Schwarzschild metric in string-Carroll expansion}

For completeness, we quickly recall how the near-horizon region of the Schwarzschild black hole is formulated as a string-Carroll expansion. We begin with the Schwarzchild metric in $4$ dimensions given by
\begin{equation}
	ds^2 = -f(r)dt^2 + \frac{dr^2}{f(r)} + r^2 (d\theta^2 + \sin^2\theta d\phi^2), \quad f(r) = 1- \frac{r_h}{r},
\end{equation}
where $r_h = 2GM$ is the horizon radius and the speed of light in Schwarzschild spacetime is set to unity, $c=1$. In the near-horizon region, we change the radial coordinate to $r= r_h + \epsilon\frac{\mathdutchcal{r}^2}{r_h}$, where $\epsilon$ is positive and dimensionless. It serves as a near-horizon expansion parameter, which goes to zero as we approach the horizon and characterises the near-horizon region for $0<\epsilon\ll 1$. Then, the Schwarzschild metric, in the near-horizon region, takes the form
\begin{equation}
	ds^2 = r_h^2 (d\theta^2 + \sin^2\theta d\phi^2) + \epsilon\Big[ -\frac{\mathdutchcal{r}^2}{r_h^2}dt^2 + 4d\mathdutchcal{r}^2 + 2\mathdutchcal{r}^2(d\theta^2 + \sin^2\theta d\phi^2)\Big] + \calO(\epsilon^2). \label{eq:nh-exp Sch}
\end{equation}
%\begin{eqnarray}
%	ds^2 &=& r_h^2 (d\theta^2 + \sin^2\theta d\phi^2) + \epsilon\Big[ -\frac{\mathdutchcal{r}^2}{r_h^2}dt^2 + 4d\mathdutchcal{r}^2 + 2\mathdutchcal{r}^2(d\theta^2 + \sin^2\theta d\phi^2)\Big] \nonumber \\
%	&& + \epsilon^2\Big[ \frac{\mathdutchcal{r}^4}{r_h^4}dt^2 + 4\frac{\mathdutchcal{r}^2}{r_h^2}d\mathdutchcal{r}^2 + \frac{\mathdutchcal{r}^4}{r_h^2} (d\theta^2 + \sin^2\theta d\phi^2)\Big] + \calO(\epsilon^3).
%\end{eqnarray}
Comparing the near-horizon expansion of the Schwarzschild metric with \eqref{eq:string-Carroll expansion}, we see that it is essentially a string-Carroll expansion with the identification $\epsilon\sim c^2$,
\begin{equation}
	ds^2 = h_{\mu\nu}dX^{\mu}dX^{\nu} + \epsilon\big(\tau_{\mu\nu} + h_{(2)\mu\nu}\big) dX^{\mu}dX^{\nu}  + \calO(\epsilon^2), \label{eq:nh-epsilon exp Sch}
\end{equation}
where the LO degenerate metrics correspond to
\begin{equation}
	h_{\mu\nu}(X)dX^{\mu}dX^{\nu} = r_h^2 (d\theta^2 + \sin^2\theta d\phi^2), \qquad \tau_{\mu\nu}(X)dX^{\mu}dX^{\nu} = -\frac{\mathdutchcal{r}^2}{r_h^2}dt^2 + 4d\mathdutchcal{r}^2. \label{eq:Rindler times sphere}
\end{equation}
%where $X^{\mu} = (t,\mathdutchcal{r},\theta,\phi)$ denote the Schwarzschild coordinates.
These describe a string-Carroll geometry, Rindler$_2\times S^2$, with $\tau_{ab}$ the metric for the longitudinal $(1+1)$-dimensional Rindler spacetime and $h_{ij}$ the metric for the transverse $2$-sphere of radius $r_h$.
%\begin{eqnarray}
%	&& h_{\mu\nu}(X)dX^{\mu}dX^{\nu} = r_h^2 (d\theta^2 + \sin^2\theta d\phi^2), \quad \tau_{\mu\nu}(X)dX^{\mu}dX^{\nu} = -\frac{\mathdutchcal{r}^2}{r_h^2}dt^2 + 4d\mathdutchcal{r}^2, \\ && h_{(2)\mu\nu}(X)dX^{\mu}dX^{\nu} = 2\mathdutchcal{r}^2(d\theta^2 + \sin^2\theta d\phi^2), \quad \tau_{(2)\mu\nu}(X)dX^{\mu}dX^{\nu} = \frac{\mathdutchcal{r}^4}{r_h^4}dt^2 + 4\frac{\mathdutchcal{r}^2}{r_h^2}d\mathdutchcal{r}^2.
%\end{eqnarray}

\subsection{The motion of the chiral Carroll string}

As noted in \cite{Bagchi:2024rje}, the identification $\epsilon\sim c^2$ holds only for the background metric. We cannot simply replace $c^2$ with $\epsilon$ in the Lagrangians because factors of $c$ appear for dimensional reasons and vanish once we set $c=1$. Thus, for clarity, we rewrite the relevant expressions from section~\ref{sec:chiral string from expansion} in the derivation  of the chiral Carroll string Lagrangian from the phase space Lagrangian of the relativistic string while keeping track of the factors of $\epsilon$.

We consider a relativistic closed bosonic string approaching the horizon of a Schwarzsch-ild black hole, as viewed by a stationary observer at infinity. The near-horizon Schwarzschild coordinates $X^{\mu} = (t,\mathdutchcal{r},\theta,\phi)$ become embedding fields $X^{\mu}(\tau,\sigma)$ for $\mu=t,\mathdutchcal{r},\theta,\phi$ 
%$= \big( X^t(\sigma^{\hat{\alpha}}), X^{\mathdutchcal{r}}(\sigma^{\hat{\alpha}}), X^{\theta}(\sigma^{\hat{\alpha}}), X^{\phi}(\sigma^{\hat{\alpha}})\big)$
that describe the embedding of the string worldsheet into the near-horizon region as the 	target spacetime. The phase space Lagrangian for the relativistic closed bosonic string with $c=1$ is given by
\begin{equation}
	{\cal L} = P_{\mu}\dot{X}^{\mu} - \frac{e}{2}\big( g^{\mu\nu}P_{\mu}P_{\nu} + T^2 g_{\mu\nu}X'^{\mu}X'^{\nu}\big) - u P_{\mu}X'^{\mu}, \label{eq:phase space L v2}
\end{equation}
where $g_{\mu\nu}$ is the Schwarzschild metric. The equation of motion for $P_{\mu}$ gives
\begin{equation}
	P_{\mu} = \frac{g_{\mu\nu}}{e}\big(\dot{X}^{\nu} - u X'^{\nu}\big).
\end{equation}
In the near-horizon region, the pre-Carrollian decomposition of the Schwarzschild metric in terms of $\epsilon$ is given by $g_{\mu\nu} = \epsilon T_{\mu\nu} + \Pi_{\mu\nu}$, which gives the longitudinal and transverse momenta
\begin{equation}
	P_{\mu}^{||} = \epsilon \frac{T_{\mu\nu}}{e}\big(\dot{X}^{\nu} - u X'^{\nu}\big), \quad P_{\mu}^{\perp} = \frac{\Pi_{\mu\nu}}{e}\big(\dot{X}^{\nu} - u X'^{\nu}\big).
\end{equation}
We define the chiral field $\psi_{\hat{\alpha}}^{\mu}$ and the anti-chiral field $\chi_{\hat{\alpha}}^{\mu}$ on the worldsheet,
\begin{equation}
	\psi_{\hat{\alpha}}^{\mu} = \frac{\gamma_{\hat{\alpha}\hat{\rho}}\Gamma^{\hat{\rho}\hat{\lambda}}_{-}\partial_{\hat{\lambda}}X^{\nu}\Pi^{\mu}_{\nu}}{\epsilon}, \quad \chi_{\hat{\alpha}}^{\mu} = \frac{\gamma_{\hat{\alpha}\hat{\rho}}\Gamma^{\hat{\rho}\hat{\lambda}}_{+}\partial_{\hat{\lambda}}X^{\nu}\Pi^{\mu}_{\nu}}{\epsilon^2},
\end{equation}
using which the transverse momentum can be written as
\begin{equation}
	P_{\mu}^{\perp} = -\frac{\Pi_{\mu\nu}}{e}\gamma^{\tau\hat{\alpha}}\big(\epsilon\psi_{\hat{\alpha}}^{\nu} + \epsilon^2 \chi_{\hat{\alpha}}^{\nu}\big). \label{eq:trans P epsilon decomposition}
\end{equation}
Integrating out the longitudinal momentum $P_{\mu}^{||}$ and using the expression for $P_{\mu}^{\perp}$ in terms of $\psi_{\hat{\alpha}}^{\mu}$, $\chi_{\hat{\alpha}}^{\mu}$, the Lagrangian \eqref{eq:phase space L v2} becomes
\begin{eqnarray}
	{\cal L}^{(c)} &=& -\frac{T_{\eff}}{2}\sqrt{-\gamma}\big(\gamma^{\hat{\alpha}\hat{\beta}}\partial_{\hat{\alpha}}X^{\mu}\partial_{\hat{\beta}}X^{\nu}T_{\mu\nu} + 2\Gamma_{+}^{\hat{\alpha}\hat{\beta}}\psi_{\hat{\alpha}}^{\mu}\partial_{\hat{\beta}}X^{\nu}\Pi_{\mu\nu} \nonumber \\
	&& + 2\epsilon \Gamma_{-}^{\hat{\alpha}\hat{\beta}}\chi_{\hat{\alpha}}^{\mu}\partial_{\hat{\beta}}X^{\nu}\Pi_{\mu\nu} - 2\epsilon^2 \gamma^{\hat{\alpha}\hat{\beta}}\psi_{\hat{\alpha}}^{\mu}\chi_{\hat{\beta}}^{\nu}\Pi_{\mu\nu}\big). \label{eq:pre-Carrollian chiral L v2}
\end{eqnarray}
where, now, we have the rescaled the tension $T$ and the Lagrange multiplier $e$ as $T_{\eff} = \epsilon T$ and $e_{\eff} =  \frac{e}{\epsilon}$, which keep the product $e_{\eff}T_{\eff} = eT$ unaffected so that the Lorentzian fiducial metric on the relativistic worldsheet is still given by \eqref{eq:rel fiducial metric v1}. The $\epsilon$-expansion of $X^{\mu}$, $\gamma_{\hat{\alpha}\hat{\beta}}$, $\psi_{\hat{\alpha}}^{\mu}$, $\chi_{\hat{\alpha}}^{\mu}$ and $\Gamma_{\pm}^{\hat{\alpha}\hat{\beta}}$ is identical to the Carroll expansion in section~\ref{sec:chiral string from expansion} with $c^2$ replaced by $\epsilon$. Then, expanding the pre-Carrollian Lagrangian ${\cal L}^{(c)}$ in powers of $\epsilon$, we get ${\cal L}^{(c)} = {\cal L}^{(c)}_{LO} + \epsilon {\cal L}^{(c)}_{NLO} + {\cal O}(\epsilon^2)$, where ${\cal L}^{(c)}_{LO}$ and ${\cal L}^{(c)}_{NLO}$ are given in \eqref{eq:chiral L LO} and \eqref{eq:chiral L NLO}, respectively, but with $T_{\eff} = \epsilon T$.

\subsubsection*{The LO equations of motion}

%\begin{equation}
%	{\cal L}^{(c)}_{LO} =  -\frac{T_{\eff}}{2}\sqrt{-\gamma_{(0)}}\big(\gamma_{(0)}^{\hat{\alpha}\hat{\beta}}\partial_{\hat{\alpha}}x^{\mu}\partial_{\hat{\beta}}x^{\nu}\tau_{\mu\nu} + 2\Gamma_{(0)+}^{\hat{\alpha}\hat{\beta}}\psi_{(0)\hat{\alpha}}^{\mu}\partial_{\hat{\beta}}x^{\nu}h_{\mu\nu}\big),
%\end{equation}
%and
%\begin{eqnarray}
%	{\cal L}^{(c)}_{NLO} &=& -\frac{T_{\eff}}{2}\sqrt{-\gamma_{(0)}}\Big[\gamma_{(0)}^{\hat{\alpha}\hat{\beta}}\big(2\partial_{\hat{\alpha}}x^{\mu}\partial_{\hat{\beta}}y^{\nu}\tau_{\mu\nu} + \partial_{\hat{\alpha}}x^{\mu}\partial_{\hat{\beta}}x^{\nu}\tau_{(2)\mu\nu}(x,y)\big) \nonumber \\
%	&& + 2\Gamma_{(0)+}^{\hat{\alpha}\hat{\beta}}\big(\psi_{(2)\hat{\alpha}}^{\mu}\partial_{\hat{\beta}}x^{\nu}h_{\mu\nu} + \psi_{(0)\hat{\alpha}}^{\mu}\partial_{\hat{\beta}}y^{\nu}h_{\mu\nu} + \psi_{(0)\hat{\alpha}}^{\mu}\partial_{\hat{\beta}}x^{\nu}h_{(2)\mu\nu}(x,y)\big) \nonumber \\
%	&& + 2\Gamma_{(0)-}^{\hat{\alpha}\hat{\beta}} \chi_{(0)\hat{\alpha}}^{\mu}\partial_{\hat{\beta}}x^{\nu}h_{\mu\nu} - \half G_{(0)}^{\hat{\alpha}\hat{\beta}\hat{\rho}\hat{\lambda}}\gamma_{(2)\hat{\rho}\hat{\lambda}}\big(\partial_{\hat{\alpha}}x^{\mu}\partial_{\hat{\beta}}x^{\nu}\tau_{\mu\nu} + \psi_{(0)\hat{\alpha}}^{\mu}\partial_{\hat{\beta}}x^{\nu}h_{\mu\nu}\big)\Big].
%\end{eqnarray}
The equations of motion for the chiral Carroll string can be obtained from the leading order Lagrangian ${\cal L}_{LO}^{(c)}$ as follows. The chiral field $\psi_{(0)\hat{\alpha}}^{\mu}$ is an auxiliary field and plays the role of a Lagrange multiplier. Thus, varying ${\cal L}_{LO}^{(C)}$ with respect to $\psi_{(0)\hat{\alpha}}^{\mu}$ gives the constraint
\begin{equation}
	\Gamma_{(0)+}^{\hat{\alpha}\hat{\beta}} \partial_{\hat{\beta}}x^{\nu}h_{\nu\mu} = 0. \label{eq:psi eom gen}
\end{equation}
%For the Schwarzschild background, $h_{ij}$ is the (invertible) sphere metric on the horizon. Thus, the constraint reduces to
%\begin{equation}
%	\Gamma_{(0)+}^{\hat{\alpha}\hat{\beta}} \partial_{\hat{\beta}}x^{i} = 0.
%\end{equation}
Varying ${\cal L}_{LO}^{(c)}$ with respect to the LO embedding field $x^{\mu}$ gives the equations of motion
\begin{eqnarray}
	&& \partial_{\hat{\alpha}}\big[\sqrt{-\gamma_{(0)}}\big(\gamma_{(0)}^{\hat{\alpha}\hat{\beta}}\partial_{\hat{\beta}}x^{\nu}\tau_{\nu\mu} + \Gamma^{\hat{\beta}\hat{\alpha}}_{{(0)}+} \psi_{{(0)}\hat{\beta}}^{\nu}h_{\nu\mu}\big)\big] \nonumber \\
	&& - \frac{\sqrt{-\gamma_{(0)}}}{2}\big[\gamma_{(0)}^{\hat{\alpha}\hat{\beta}}\partial_{\hat{\alpha}}x^{\rho}\partial_{\hat{\beta}}x^{\lambda}\partial_{\mu}\tau_{\rho\lambda} + 2\Gamma_{{(0)}+}^{\hat{\alpha}\hat{\beta}} \psi_{{(0)}\hat{\alpha}}^{\rho}\partial_{\hat{\beta}}x^{\lambda}\partial_{\mu}h_{\rho\lambda}\big] = 0. \label{eq:x mu eom gen}
\end{eqnarray}
Finally varying ${\cal L}_{LO}^{(C)}$ with respect to the LO worldsheet metric $\gamma_{(0)}^{\hat{\alpha}\hat{\beta}}$ gives the LO Virasoro constraints $T_{(0)\hat{\alpha}\hat{\beta}} = 0$, where the worldsheet LO stress tensor $T_{(0)\hat{\alpha}\hat{\beta}}$ is given by
\begin{equation}
	T_{{(0)}\hat{\alpha}\hat{\beta}} = \partial_{\hat{\alpha}}x^{\mu}\partial_{\hat{\beta}}x^{\nu}\tau_{\mu\nu} + \psi_{(0)(\hat{\alpha}}^{\mu}\partial_{\hat{\beta})}x^{\nu}h_{\mu\nu} - \frac{\gamma_{(0)\hat{\alpha}\hat{\beta}}}{2}\gamma_{(0)}^{\hat{\rho}\hat{\lambda}}\big(\partial_{\hat{\rho}}x^{\mu}\partial_{\hat{\lambda}}x^{\nu}\tau_{\mu\nu} + \psi_{(0)\hat{\rho}}^{\mu}\partial_{\hat{\lambda}}x^{\nu}h_{\mu\nu}\big) \label{eq:Vir constraint gen}
\end{equation}
using the definition, $T_{{(0)}\hat{\alpha}\hat{\beta}} \equiv -\frac{2}{T_{\eff}\sqrt{-\gamma_{(0)}}} \frac{\delta S^{(C)}_{LO}}{\delta \gamma_{(0)}^{\hat{\alpha}\hat{\beta}}}$.
%\begin{align}
%	\hspace{-3mm} T_{{(0)}\hat{\alpha}\hat{\beta}} &\equiv -\frac{2}{T_{\eff}\sqrt{-\gamma_{(0)}}} \frac{\delta S^{(C)}_{LO}}{\delta \gamma_{(0)}^{\hat{\alpha}\hat{\beta}}} \nonumber \\
%	&= \partial_{\hat{\alpha}}x^{\mu}\partial_{\hat{\beta}}x^{\nu}\tau_{\mu\nu} + \psi_{(0)(\hat{\alpha}}^{\mu}\partial_{\hat{\beta})}x^{\nu}h_{\mu\nu} - \frac{\gamma_{(0)\hat{\alpha}\hat{\beta}}}{2}\gamma_{(0)}^{\hat{\rho}\hat{\lambda}}\big(\partial_{\hat{\rho}}x^{\mu}\partial_{\hat{\lambda}}x^{\nu}\tau_{\mu\nu} + \psi_{(0)\hat{\rho}}^{\mu}\partial_{\hat{\lambda}}x^{\nu}h_{\mu\nu}\big).
%\end{align}

\subsubsection*{Solution to LO equations of motion on flat worldsheet}

The chiral Carroll string has a Lorentzian worldsheet, and the action $S_{LO}^{(c)} = \int d^2\sigma {\cal L}_{LO}^{(c)}$ is invariant under worldsheet diffeomorphisms, $\sigma^{\hat{\alpha}}\rightarrow \tilde{\sigma}^{\hat{\alpha}}(\sigma^{\hat{\beta}})$ and local Weyl rescaling, $\gamma_{(0)\hat{\alpha}\hat{\beta}}\rightarrow e^{2\omega}\gamma_{(0)\hat{\alpha}\hat{\beta}}$. Using these local worldsheet symmetries, we gauge fix the worldsheet metric to be the Minkowski metric, $\gamma_{(0)\hat{\alpha}\hat{\beta}} = \eta_{\hat{\alpha}\hat{\beta}}$.
%\footnote{A detailed discussion on the worldsheet symmetries and their gauge fixing can be found in \cite{Bagchi:2023cfp}.}
%\qquad \gamma_{(2)\hat{\alpha}\hat{\beta}} = 0.
On the flat worldsheet in Cartesian coordinates, we have $\sqrt{-\gamma_{(0)}} = 1$ giving $\varepsilon_{(0)}^{\hat{\alpha}\hat{\beta}} = \boldsymbol{\varepsilon}_{(0)}^{\hat{\alpha}\hat{\beta}}$, in particular $\varepsilon_{(0)}^{\tau\sigma} = \boldsymbol{\varepsilon}_{(0)}^{\tau\sigma} = 1$. Now, anticipating the chiral nature of the chiral Carroll string, we change the worldsheet coordinates to lightcone coordinates, $\sigma^{\pm} = \tau \pm \sigma$. In lightcone coordinates, the components of the antisymmetric symbol are given by $\varepsilon_{(0)}^{+-} = -2 = - \varepsilon_{(0)}^{-+}$ and the worldsheet metric is
\begin{equation}
	\gamma_{(0)\hat{\alpha}\hat{\beta}} = \begin{pmatrix}
		0 & -\frac{1}{2} \\ -\frac{1}{2} & 0
	\end{pmatrix}, \qquad \gamma_{(0)}^{\hat{\alpha}\hat{\beta}} = \begin{pmatrix}
	0 & -2 \\ -2 & 0
	\end{pmatrix}, \qquad \sqrt{-\gamma_{(0)}} = \frac{1}{2},
\end{equation}
which together give
%\begin{equation}
%	\varepsilon_{(0)}^{+-} = \frac{\partial \sigma^+}{\partial \tau}\frac{\partial \sigma^-}{\partial \sigma}\varepsilon_{(0)}^{\tau\sigma} + \frac{\partial \sigma^+}{\partial \sigma}\frac{\partial \sigma^-}{\partial \tau}\varepsilon_{(0)}^{\sigma\tau} = -2, \qquad \varepsilon_{(0)}^{-+} = 2,
%\end{equation}
the components of the chiral projectors,
\begin{equation}
	\Gamma_{(0)+}^{+-} = -2, \quad \Gamma_{(0)+}^{-+} = 0, \quad \Gamma_{(0)-}^{+-} = 0, \quad \Gamma_{(0)-}^{-+} = -2.
\end{equation}

Now, we solve the LO equations of motion one by one on the Minkowski worldsheet and in lightcone coordinates as follows. From \eqref{eq:Rindler times sphere}, using $h_{\theta\theta} = r_h^2$, $h_{\phi\phi} = r_h^2\sin^2x^{\theta}$, the equation of motion \eqref{eq:psi eom gen} for $\psi_{(0)\hat{\alpha}}^{\mu}$ reduces to $\partial_-x^{i} = 0$, yielding
\begin{equation}
	x^{\theta} = x^{\theta}(\sigma^+), \quad x^{\phi} = x^{\phi}(\sigma^+),
\end{equation}
\ie~the transverse LO embedding fields $x^i$ are chiral functions of $\sigma^+$. The equation of motion \eqref{eq:x mu eom gen} for $\mu = \theta,\phi$ reduces to $\partial_-\psi_{(0)+}^i = 0$,
%\begin{equation}
%	\partial_-\psi_{(0)+}^{\theta} = 0, \quad \sin^2 x^{\theta}\partial_-\psi_{(0)+}^{\phi} = 0,
%\end{equation}
which implies that the auxiliary fields are also chiral functions of $\sigma^{+}$ on the worldsheet,
\begin{equation}
	\psi_{(0)+}^{\theta} = \psi_{(0)+}^{\theta}(\sigma^+), \quad \psi_{(0)+}^{\phi} = \psi_{(0)+}^{\phi}(\sigma^+).
\end{equation}
We also note that although $\psi_{(0)-}^{\mu}$ does not appear dynamically, we get that $\psi_{(0)-}^{\mu} = 0$ from the definition itself, \ie~from the LO term in the Carroll expansion of \eqref{eq:psi-psi-equivalence}. Then, from the leading term in the $\epsilon$-expansion of \eqref{eq:trans P epsilon decomposition}, we get that the transverse conjugate momentum is chiral, \ie~$P^{\perp}_{(0)i} = P^{\perp}_{(0)i}(\sigma^+)$ for $i=\theta,\phi$.

The equation of motion \eqref{eq:x mu eom gen} for $\mu = t,\ \mathdutchcal{r}$ reduces to
\begin{subequations}\label{eq:x-t-rho eom Sch}
	\begin{align}
		(x^{\mathdutchcal{r}})^2\partial_+\partial_-x^t + x^{\mathdutchcal{r}}(\partial_+x^t\partial_-x^{\mathdutchcal{r}} + \partial_-x^t\partial_+x^{\mathdutchcal{r}}) &= 0, \label{eq:xt eom Sch} \\
		4\partial_+\partial_-x^{\mathdutchcal{r}} + \frac{x^{\mathdutchcal{r}}}{r_h^2}\partial_+x^t\partial_-x^t &= 0, \label{eq:xrho eom Sch}
	\end{align}
\end{subequations}
and the LO Virasoro constraints \eqref{eq:Vir constraint gen} reduce to
\begin{subequations}
	\begin{align}
		-\frac{(x^{\mathdutchcal{r}})^2}{r_h^2}(\partial_+x^t)^2 + 4(\partial_+x^{\mathdutchcal{r}})^2 + r_h^2 (\psi_{(0)+}^{\theta}\partial_+x^{\theta} + \sin^2 x^{\theta}\psi_{(0)+}^{\phi}\partial_+ x^{\phi}) &= 0, \label{eq:++Vir Sch}\\
		-\frac{(x^{\mathdutchcal{r}})^2}{r_h^2}(\partial_- x^t)^2 + 4(\partial_- x^{\mathdutchcal{r}})^2 &= 0. \label{eq:--Vir Sch}
	\end{align}
\end{subequations}
%
%
%
%\begin{eqnarray}
%	&& 2\partial_+\partial_-x^{\nu}\tau_{\nu\mu} + \partial_+x^{\nu}\partial_-\tau_{\nu\mu} + \partial_-x^{\nu}\partial_+\tau_{\nu\mu}  - \partial_+x^{\mathdutchcal{r}}\partial_-x^{\lambda}\partial_{\mu}\tau_{\mathdutchcal{r}\lambda} \nonumber \\
%	&& + \partial_-(\ost{\psi}{\zero}_+^{\nu}h_{\nu\mu}) - \ost{\psi}{\zero}_+^{\mathdutchcal{r}}\partial_-x^{\lambda}\partial_{\mu}h_{\mathdutchcal{r}\lambda} = 0.
%\end{eqnarray}
%
%The LO Virasoro constraints reduce to
%\begin{equation}
%	\ost{T}{\zero}_{\pm\pm} = \partial_{\pm}x^{\mu}\partial_{\pm}x^{\nu}\tau_{\mu\nu} + \ost{\psi}{\zero}_{\pm}^{\mu}\partial_{\pm}x^{\nu}h_{\mu\nu} = 0.
%\end{equation}
%Now using $\partial_-x^{\nu}h_{\mu\nu} = 0$, the LO Virasoro constraints further simplify to
%\begin{eqnarray}
%	\ost{T}{\zero}_{++} &=& \partial_{+}x^{\mu}\partial_{+}x^{\nu}\tau_{\mu\nu} + \ost{\psi}{\zero}_{+}^{\mu}\partial_{+}x^{\nu}h_{\mu\nu} = 0. \\
%	\ost{T}{\zero}_{--} &=& \partial_{-}x^{\mu}\partial_{-}x^{\nu}\tau_{\mu\nu} = 0.
%\end{eqnarray}
%
%
%
%\subsubsection*{Solution to LO equations of motion}
%
To find the solutions for $x^t$ and $x^{\mathdutchcal{r}}$, it is convenient to go to the flat lightcone coordinates in the Rindler spacetime, defined by
\begin{equation}
	x^+ = 2x^{\mathdutchcal{r}} e^{\frac{x^t}{2r_h}}, \quad x^- = - 2x^{\mathdutchcal{r}} e^{-\frac{x^t}{2r_h}}, \label{eq:t rho to pm transf}
\end{equation}
such that the Rindler metric becomes the flat metric
\begin{equation}
	ds^2_{Rindler} = -\frac{(x^{\mathdutchcal{r}})^2}{r_h^2}(dx^t)^2 + 4(dx^{\mathdutchcal{r}})^2 = -dx^+ dx^-.
\end{equation}
In these flat coordinates, the equations of motion \eqref{eq:x-t-rho eom Sch} take the form of the wave equations
\begin{equation}
	\partial_+ \partial_- x^+ = 0, \quad \partial_+ \partial_- x^- = 0,
\end{equation}
whose general solution can be written as
\begin{equation}
	x^+ = x^+_L(\sigma^+) + x^+_R(\sigma^-), \quad x^- = x^-_L(\sigma^+) + x^-_R(\sigma^-).
\end{equation}
Substituting these solutions in the Virasoro constraint $T_{(0)--} = 0$ in \eqref{eq:--Vir Sch}, it reduces to
\begin{equation}
	\partial_- x^+_R \partial _- x^-_R = 0 \quad \implies \quad x^+_R(\sigma^-) = 0\quad \text{or}\quad  x^-_R(\sigma^-) = 0.
\end{equation}
The remaining Virasoro constraint $T_{(0)++} = 0$ in \eqref{eq:++Vir Sch} then constraints the left moving modes of $x^{\pm}$ as
\begin{equation}
	\partial_+x^+_L\partial_+x^-_L = r_h^2 (\psi_{(0)+}^{\theta}\partial_+x^{\theta} + \sin^2 x^{\theta}\psi_{(0)+}^{\phi}\partial_+ x^{\phi}). \label{eq:T++ constraint onshell}
\end{equation}
Then, to write the solution for $x^t$ and $x^{\mathdutchcal{r}}$, we can simply do the inverse transformation of \eqref{eq:t rho to pm transf}.

%In summary, the most general motion of the chiral Carroll string as it approaches the horizon of the Schwarzschild black hole is given by \textcolor{red}{to be continued from here}.

\subsection{Comparison with the magnetic Carroll string}

%The chiral and magnetic Carroll strings, both, have Lorentzian worldsheets, and
In the preceding subsection, we found that the chiral Carroll string has left-moving fluctuations on the horizon given by $x^{\theta}(\sigma^+)$, $x^{\phi}(\sigma^+)$, whereas the magnetic Carroll string freezes on the horizon, \ie~$x^{\theta},x^{\phi}\sim$ constant \cite{Bagchi:2024rje}.
%that the magnetic Carroll string has a Lorentzian worldsheet and shrinks to a point on the transverse sphere, \ie~$x^{\theta}\sim$ constant and $x^{\phi}\sim$ constant. It moves along a null geodesic as a point particle or a folded string in the longitudinal Rindler space. The chiral Carroll string also has a Lorentzian worldsheet but has a right-moving chiral motion for the transverse sphere coordinates, $x^{\theta}(\sigma^+)$ and $x^{\phi}(\sigma^+)$.
Comparing the equations of motion and the Virasoro constraints for $x^t$, $x^{\mathdutchcal{r}}$ for the chiral and magnetic Carroll strings, we see that the only difference arises in the $T_{(0)++}=0$ Virasoro constraint \eqref{eq:++Vir Sch} or \eqref{eq:T++ constraint onshell}. Then, if we further set $x^{\theta}(\sigma^+)$, $x^{\phi}(\sigma^+)$ to be constants, which is also a solution for the chiral Carroll string, the constraint \eqref{eq:T++ constraint onshell} reduces to that of the magnetic Carroll string. Thus, we see that the magnetic Carroll string arises as a special case of the chiral Carroll string.

Having compared the dynamics, we now look at the boundary conditions on the horizon for the chiral and magnetic Carroll strings. In \cite{Bagchi:2024rje}, the magnetic boundary conditions on the horizon are derived from the leading order terms in the $\epsilon$-expansion of the Hamiltonian constraints (that are linear combinations of the Virasoro constraints), which in lightcone coordinates are given by
\begin{equation}
	\text{Magnetic}: \qquad \partial_{+}X^i\vert_{r_h} = 0, \quad \partial_{-}X^i\vert_{r_h} = 0; \quad \tilde{e}\tilde{T} = 1.
\end{equation}
However, looking at \eqref{eq:++Vir Sch}, \eqref{eq:--Vir Sch}, the non-triviality of the leading order terms in the $\epsilon$-expansion of the Virasoro constraints for the chiral Carroll string allows us to set $\partial_- X^i = 0$ but requires $\partial_+ X^i \neq 0$ on the horizon, thus, giving the chiral boundary conditions
\begin{equation}
	\text{Chiral}: \qquad \partial_{-}X^i\vert_{r_h} = 0, \quad \partial_{+}X^i\vert_{r_h} \neq 0; \quad e_{\eff}T_{\eff} = 1
\end{equation}
on the horizon.

%%%%%%%%%%%%%%%%%%%%%%%%%%%
%%%%%% Discussion %%%%%%%%%%%%%%%

\section{Discussion}\label{sec:discussion}

In summary, we have presented an alternative formulation for the magnetic Carroll expansion of the classical worldsheet action for the relativistic closed bosonic string. We did so by doing the Carroll expansion of a Polyakov form of the relativistic string action which contains the unintegrated transverse momentum expressed in terms of an auxiliary field on the worldsheet. Then, decomposing the transverse momentum into its chiral and anti-chiral parts through chiral and anti-chiral auxiliary fields on the worldsheet but with a relative factor of $c^2$ between them, we defined the chiral Carroll expansion of the relativistic string action. The leading order term in this expansion gives the action for the chiral Carroll string moving in a curved string-Carroll target spacetime.

We then studied the dynamics of a classical relativistic closed bosonic string approaching the Schwarzschild black hole, when it behaves as the chiral Carroll string moving in the near-horizon string-Carroll expanded geometry with a $2$-dimensional longitudinal Rinder spacetime and a $2$-dimensional transverse sphere. We find that the transverse embedding fields, \ie~the horizon coordinates have arbitrary dependence on only the left-moving lightcone coordinate on the worldsheet. In the longitudinal Rindler spacetime expressed in flat lightcone coordinates, one lightcone embedding field is also chiral depending arbitrarily only on the left-moving lightcone coordinate, whereas the solution for the remaining longitudinal embedding field depends on the transverse fields through one of the Virasoro constraints.

In this paper, we have written down the most general solution for the motion of the chiral Carroll string moving in the near-horizon region of a Schwarzschild black hole. It would be interesting to find particular solutions for specific forms of the arbitrary functions $x^{\theta}(\sigma^+)$, $x^{\phi}(\sigma^+)$, $x^+_L(\sigma^+)$ (or $x^-_L(\sigma^+)$) and $\psi_{(0)+}^{\theta}(\sigma^+)$, $\psi_{(0)+}^{\phi}(\sigma^+)$, and their physical interpretation for the string's motion in the region near a black hole horizon. In particular, it would be interesting to explore how the magnetic Carroll string's behaviour in the Rindler spacetime $-$ as a massless particle and a folded string, elaborated on in \cite{Bagchi:2024rje,Banerjee:2025bkg} $-$ deforms when small chiral perturbations of the transverse horizon coordinates are turned on.

\section*{Acknowledgements}

We thank Arjun Bagchi for useful discussions. We would also like to thank Arjun Bagchi, Aritra Banerjee, Jelle Hartong, Emil Have and Mangesh Mandlik for collaboration on our earlier work on Carroll strings. Some of the computations in Appendix~\ref{app:magnetic string near black hole} are performed with the help of an AI assistant. This work was presented at the International Congress of Basic Science (ICBS) 2026 held at Beijing Institute of Mathematical Sciences and Applications, Beijing, and we thank the organisers for their hospitality and the participants for interesting discussions.

\appendix

\section{Magnetic Carroll string near Schwarzschild black hole}\label{app:magnetic string near black hole}

In this appendix, we explicitly show the equivalence of the two descriptions of the magnetic Carroll string, given in \cite{Bagchi:2024rje} and in section~\ref{sec:magnetic string}, by verifying the equations of motion for the LO and NLO embedding fields $x^{\mu}(\tau,\sigma)$ and $y^{\mu}(\tau,\sigma)$ in the near-horizon region of the Schwarzschild black hole.

The near-horizon expansion of the Schwarzschild metric \eqref{eq:nh-exp Sch}, now including ${\cal O}(\epsilon^2)$ terms, is given by
\begin{align}
	ds^2 &= r_h^2 (d\theta^2 + \sin^2\theta d\phi^2) + \epsilon\Big[ -\frac{\mathdutchcal{r}^2}{r_h^2}dt^2 + 4d\mathdutchcal{r}^2 + 2\mathdutchcal{r}^2(d\theta^2 + \sin^2\theta d\phi^2)\Big] \nonumber \\
	& \quad + \epsilon^2\Big[ \frac{\mathdutchcal{r}^4}{r_h^4}dt^2 + 4\frac{\mathdutchcal{r}^2}{r_h^2}d\mathdutchcal{r}^2 + \frac{\mathdutchcal{r}^4}{r_h^2} (d\theta^2 + \sin^2\theta d\phi^2)\Big] + \calO(\epsilon^3).
\end{align}
Comparing with the string-Carroll expansion \eqref{eq:string-Carroll expansion} with the identification $\epsilon\sim c^2$,
\begin{equation}
	ds^2 = \big[h_{\mu\nu} + \epsilon\big(\tau_{\mu\nu} + h_{(2)\mu\nu}\big) + \epsilon^2 \big(\tau_{(2)\mu\nu} + h_{(4)\mu\nu}\big)\big]dX^{\mu}dX^{\nu}  + \calO(\epsilon^3),
\end{equation}
gives the LO degenerate metrics \eqref{eq:Rindler times sphere} and the NLO geometric fields
\begin{equation}
	h_{(2)\mu\nu}(X)dX^{\mu}dX^{\nu} = 2\mathdutchcal{r}^2(d\theta^2 + \sin^2\theta d\phi^2), \qquad \tau_{(2)\mu\nu}(X)dX^{\mu}dX^{\nu} = \frac{\mathdutchcal{r}^4}{r_h^4}dt^2 + 4\frac{\mathdutchcal{r}^2}{r_h^2}d\mathdutchcal{r}^2.
\end{equation}
%For a relativistic closed bosonic string approaching the horizon, the $\epsilon$-expansion of the embedding fields $X^{\mu} = x^{\mu} + \epsilon y^{\mu} + {\cal O}(\epsilon^2)$ induces a Taylor expansion on the geometric fields given by
%\begin{equation}
%	\tau_{(2)\mu\nu}(x,y) \equiv \tau_{(2)\mu\nu}(x) + y^{\rho}\partial_{\rho}\tau_{\mu\nu}(x), \qquad h_{(2)\mu\nu}(x,y) \equiv h_{(2)\mu\nu}(x) + y^{\rho}\partial_{\rho}h_{\mu\nu}(x).
%\end{equation}

%\subsection*{In the description of section~\ref{sec:magnetic string}}

\medskip

The equations of motion and Virasoro constraints from the LO magnetic Lagrangian \eqref{eq:mag LO L v1} with the near-horizon geometry of the Schwarzschild black hole as a string-Carroll expanded target space, and in lightcone coordinates on flat Minkowski worldsheet, are given by
\begin{subequations}\label{eq:mixed mag LO eqns}
	\begin{align}
		\partial_{\hat{\alpha}}x^i = 0,& \hspace{20mm} \partial_+\kappa_{(0)-}^i + \partial_-\kappa_{(0)+}^i = 0, \\
		(x^{\mathdutchcal{r}})^2\partial_+\partial_-x^t + x^{\mathdutchcal{r}}(\partial_+x^t\partial_-x^{\mathdutchcal{r}} + \partial_-x^t\partial_+x^{\mathdutchcal{r}}) = 0, & \hspace{11mm} 4\partial_+\partial_-x^{\mathdutchcal{r}} + \frac{x^{\mathdutchcal{r}}}{r_h^2}\partial_+x^t\partial_-x^t = 0, \\
		-\frac{(x^{\mathdutchcal{r}})^2}{r_h^2}(\partial_+x^t)^2 + 4(\partial_+x^{\mathdutchcal{r}})^2 = 0, & \hspace{5mm} -\frac{(x^{\mathdutchcal{r}})^2}{r_h^2}(\partial_- x^t)^2 + 4(\partial_- x^{\mathdutchcal{r}})^2 = 0.
	\end{align}
\end{subequations}
The equations of motion and Virasoro constraints from the NLO magnetic Lagrangian \eqref{eq:mag L NLO v1}, upon using the LO equations \eqref{eq:mixed mag LO eqns}, simplify to
\begin{subequations}\label{eq:mixed mag NLO eqns}
\begin{align}
	& \partial_{\hat{\alpha}}y^i = \kappa_{(0)\hat{\alpha}}^i \quad \implies \qquad \partial_+\partial_- y^i = 0, \\
	& \partial_+\Big[\frac{(x^{\mathdutchcal{r}})^2}{r_h^2}\partial_- y^t - \Big(\frac{(x^{\mathdutchcal{r}})^4}{r_h^4} - \frac{2x^{\mathdutchcal{r}}y^{\mathdutchcal{r}}}{r_h^2}\Big)\partial_-x^t\Big] + \partial_-\Big[\frac{(x^{\mathdutchcal{r}})^2}{r_h^2}\partial_+ y^t - \Big(\frac{(x^{\mathdutchcal{r}})^4}{r_h^4} - \frac{2x^{\mathdutchcal{r}}y^{\mathdutchcal{r}}}{r_h^2}\Big)\partial_+x^t\Big] = 0, \\
	& \partial_+\partial_-y^{\mathdutchcal{r}} + \frac{(x^{\mathdutchcal{r}})^2}{r_h^2}\partial_+\partial_-x^{\mathdutchcal{r}} + \frac{x^{\mathdutchcal{r}}}{r_h^2}\partial_+x^{\mathdutchcal{r}}\partial_-x^{\mathdutchcal{r}} + \frac{x^{\mathdutchcal{r}}}{2r_h^2}\partial_{(+}x^t\partial_{-)}y^t - \Big( \frac{(x^{\mathdutchcal{r}})^3}{2r_h^4} - \frac{y^{\mathdutchcal{r}}}{4r_h^2}\Big)\partial_+x^t\partial_-x^t = 0, \\
	& 2\frac{(x^{\mathdutchcal{r}})^2}{r_h^2}\partial_{\pm}x^t \partial_{\pm}y^t - 8\partial_{\pm}x^{\mathdutchcal{r}}\partial_{\pm}y^{\mathdutchcal{r}} - (\partial_{\pm}x^t)^2\Big(\frac{(x^{\mathdutchcal{r}})^4}{r_h^4} - \frac{x^{\mathdutchcal{r}}y^{\mathdutchcal{r}}}{r_h^2}\Big) - r_h^2\big[(\partial_{\pm}y^{\theta})^2 + \sin^2x^{\theta}(\partial_{\pm}y^{\phi})^2\big] = 0,
\end{align}
\end{subequations}
and equations of motion for $\kappa_{(2)\hat{\alpha}}^i$ that coupled with the next-to-next-to-leading order (NNLO) equations give the wave equation for $z^i$.

%\subsection*{In the description of \cite{Bagchi:2024rje}}

\medskip

In \cite{Bagchi:2024rje}, the equations of motion and Virasoro constraints from the LO and NLO Lagrangians are found to be
\begin{subequations}\label{eq:mag LO-NLO eqns}
	\begin{align}
		\partial_{\hat{\alpha}}x^i = 0,& \\
		(x^{\mathdutchcal{r}})^2\partial_+\partial_-x^t + x^{\mathdutchcal{r}}(\partial_+x^t\partial_-x^{\mathdutchcal{r}} + \partial_-x^t\partial_+x^{\mathdutchcal{r}}) = 0, &  \hspace{11mm} 4\partial_+\partial_-x^{\mathdutchcal{r}} + \frac{x^{\mathdutchcal{r}}}{r_h^2}\partial_+x^t\partial_-x^t = 0, \\
		-\frac{(x^{\mathdutchcal{r}})^2}{r_h^2}(\partial_+x^t)^2 + 4(\partial_+x^{\mathdutchcal{r}})^2 = 0, & \hspace{5mm} -\frac{(x^{\mathdutchcal{r}})^2}{r_h^2}(\partial_- x^t)^2 + 4(\partial_- x^{\mathdutchcal{r}})^2 = 0, \\
		\partial_+\partial_- y^i = 0.&
	\end{align}
\end{subequations}
In the formulation of \cite{Bagchi:2024rje}, where the longitudinal and transverse embedding fields appear one order in $\epsilon$ apart in the Lagrangians, the equations for $y^t$ and $y^{\mathdutchcal{r}}$ come from the NNLO Lagrangian. With the help of an AI assistant, the equations of motion and Virasoro constraints involving $y^t$ and $y^{\mathdutchcal{r}}$ are computed from the NNLO Lagrangian, and simplified using the LO and NLO equations \eqref{eq:mag LO-NLO eqns} to get
\begin{subequations}\label{eq:mag NNLO eqns}
\begin{align}
	& \partial_+\Big[\frac{(x^{\mathdutchcal{r}})^2}{r_h^2}\partial_- y^t - \Big(\frac{(x^{\mathdutchcal{r}})^4}{r_h^4} - \frac{2x^{\mathdutchcal{r}}y^{\mathdutchcal{r}}}{r_h^2}\Big)\partial_-x^t\Big] + \partial_-\Big[\frac{(x^{\mathdutchcal{r}})^2}{r_h^2}\partial_+ y^t - \Big(\frac{(x^{\mathdutchcal{r}})^4}{r_h^4} - \frac{2x^{\mathdutchcal{r}}y^{\mathdutchcal{r}}}{r_h^2}\Big)\partial_+x^t\Big] = 0, \\
	& \partial_+\partial_-y^{\mathdutchcal{r}} + \frac{(x^{\mathdutchcal{r}})^2}{r_h^2}\partial_+\partial_-x^{\mathdutchcal{r}} + \frac{x^{\mathdutchcal{r}}}{r_h^2}\partial_+x^{\mathdutchcal{r}}\partial_-x^{\mathdutchcal{r}} + \frac{x^{\mathdutchcal{r}}}{2r_h^2}\partial_{(+}x^t\partial_{-)}y^t - \Big( \frac{(x^{\mathdutchcal{r}})^3}{2r_h^4} - \frac{y^{\mathdutchcal{r}}}{4r_h^2}\Big)\partial_+x^t\partial_-x^t = 0, \\
	& 2\frac{(x^{\mathdutchcal{r}})^2}{r_h^2}\partial_{\pm}x^t \partial_{\pm}y^t - 8\partial_{\pm}x^{\mathdutchcal{r}}\partial_{\pm}y^{\mathdutchcal{r}} - (\partial_{\pm}x^t)^2\Big(\frac{(x^{\mathdutchcal{r}})^4}{r_h^4} - \frac{x^{\mathdutchcal{r}}y^{\mathdutchcal{r}}}{r_h^2}\Big) - r_h^2\big[(\partial_{\pm}y^{\theta})^2 + \sin^2x^{\theta}(\partial_{\pm}y^{\phi})^2\big] = 0.
\end{align}
\end{subequations}

\medskip

We see that the equations \eqref{eq:mixed mag LO eqns}-\eqref{eq:mixed mag NLO eqns} and \eqref{eq:mag LO-NLO eqns}-\eqref{eq:mag NNLO eqns} are exactly identical, thus showing the equivalence of the two descriptions for the magnetic Carroll string dynamically.

%We see that the equations \eqref{eq:mixed mag LO eqns}-\eqref{eq:mixed mag NLO eqns}
%obtained in the mixed phase space description
%are identical to the equations \eqref{eq:mag LO-NLO eqns}-\eqref{eq:mag NNLO eqns} obtained in the Polyakov description, thus showing the equivalence of the two descriptions dynamically.

%%%%%%%%%%%%%%%%%%%%%%%%%%%
%%%%% Chiral string-Carroll symmetries %%%%%

%\section{Chiral enhancement of string-Carroll symmetries}

\addcontentsline{toc}{section}{\refname}

\bibliographystyle{JHEP}
\bibliography{Masterbibliography}

\end{document}